\documentclass[aps,pra,reprint,superscriptaddress]{revtex4-2}

\usepackage[english]{babel}

\usepackage{amsfonts,amsmath,amssymb,amsthm}
\usepackage{graphicx}

\usepackage[dvipsnames, table]{xcolor}
\usepackage{array}
\usepackage{enumitem}
\usepackage{mathpazo}
\usepackage{setspace}
\usepackage{bm}
\usepackage{amssymb}
\usepackage{amsthm}
\usepackage{mathtools}
\usepackage{multirow}
\usepackage{cases}

\usepackage[colorlinks=true,linkcolor=blue,citecolor=red,linktocpage=true,breaklinks=true]{hyperref}
\usepackage{mathptmx}
\usepackage[framemethod=default]{mdframed}
\usepackage{tcolorbox}
\usepackage{tikz} 
\usepackage{url}
\usepackage{varwidth}

\usepackage{pgfplots}

\usepackage{youngtab}
\usepackage{subfigure}
\usepackage{verbatim}
\usepackage[capitalise]{cleveref}
\usepackage{float}

\usepackage{tabularx}
\usepackage{listings}

\usepackage{tikz-3dplot}

\usetikzlibrary{positioning, calc}

\usetikzlibrary{shapes.geometric, arrows}

\usepackage{quantikz}

\begin{document}

\title{Application of Large Language Models to Quantum State Simulation
}
\author{Shuangxiang Zhou}

\affiliation{College of Physics and Electronic Engineering, Center for Computational Sciences,  Sichuan Normal University, Chengdu 610068, China}
\affiliation{Department of Physics, The Hong Kong University of Science and Technology, Clear Water Bay, Kowloon, Hong Kong, China}

\author{Ronghang Chen}

\affiliation{College of Physics and Electronic Engineering, Center for Computational Sciences,  Sichuan Normal University, Chengdu 610068, China}
\affiliation{Department of Physics, The Hong Kong University of Science and Technology, Clear Water Bay, Kowloon, Hong Kong, China}

\author{Zheng An}

\affiliation{Department of Physics, The Hong Kong University of Science and Technology, Clear Water Bay, Kowloon, Hong Kong, China}

\author{Shi-Yao Hou}

\affiliation{College of Physics and Electronic Engineering, Center for Computational Sciences,  Sichuan Normal University, Chengdu 610068, China}
\affiliation{Department of Physics, The Hong Kong University of Science and Technology, Clear Water Bay, Kowloon, Hong Kong, China}

\begin{abstract}
Quantum computers leverage the unique advantages of quantum mechanics to achieve acceleration over classical computers for certain problems. Currently, various quantum simulators provide powerful tools for researchers, but simulating quantum evolution with these simulators often incurs high time costs. Additionally, resource consumption grows exponentially as the number of quantum bits increases. To address this issue, our research aims to utilize Large Language Models (LLMs) to simulate quantum circuits. This paper details the process of constructing 1-qubit and 2-qubit quantum simulator models, extending to multiple qubits, and ultimately implementing a 3-qubit example. Our study demonstrates that LLMs can effectively learn and predict the evolution patterns among quantum bits, with minimal error compared to the theoretical output states. Even when dealing with quantum circuits comprising an exponential number of quantum gates, LLMs remain computationally efficient. Overall, our results highlight the potential of LLMs to predict the outputs of complex quantum dynamics, achieving speeds far surpassing those required to run the same process on a quantum computer. This finding provides new insights and tools for applying machine learning methods in the field of quantum computing.
\end{abstract}
\maketitle

\section{Introduction}~{}

After decades of development, quantum computers have made significant progress in several key quantum algorithms. However, from a practical perspective, they still face numerous challenges \cite{computer1,computer2}. The core component, the quantum bit (qubit), has a very short coherence time and is highly sensitive to environmental factors, making large-scale and stable manufacturing difficult \cite{quantum1}. Additionally, precisely controlling quantum gate operations is challenging, and quantum errors can accumulate easily, leading to computational failures. These issues necessitate extremely precise control technologies in constructing practical quantum computers, imposing high demands on hardware manufacturing and operating environments, which significantly increases production costs. Furthermore, maintaining quantum computers requires highly specialized skills, further elevating operational costs. Therefore, the widespread application of quantum computers across various fields faces significant challenges.

Currently, various quantum simulators have made progress in simulating basic quantum systems, such as simulating the evolution of a small number of qubits on classical computers \cite{hai2,simulation3,simulation4}. These simulators provide a foundation for understanding quantum behaviors but still present idealized scenarios and cannot comprehensively describe the complexities of quantum computers. Moreover, existing simulators exhibit high time complexity when simulating circuits, with an increase in the number of qubits leading to exponential resource consumption \cite{simulation1,computer1}.

In recent years, significant advancements have been made in the fields of machine learning and natural language processing (NLP), especially concerning the application of large language models (LLMs). For instance, in text generation tasks, LLMs like GPT-3 have demonstrated strong capabilities \cite{gpt3,llm8}. Research in sentiment analysis has shown that these models can effectively identify emotional tendencies in text \cite{language,language2}. Additionally, LLMs have achieved remarkable results in machine translation, significantly enhancing accuracy and fluency \cite{nlp3,nlp6}. These advancements not only address many complex problems but also provide powerful tools for data analysis and prediction in dynamic systems.

LLMs also exhibit potential in quantum information science, aiding in simulating quantum states and predicting the behavior of quantum systems \cite{llm2, llm3}. Their ability to handle high-dimensional data and capture complex dependencies makes them valuable for tasks such as quantum state classification and circuit analysis\cite{fenl,fenl2}. Based on these advancements, we explore the application of LLMs in simulating quantum systems, showcasing their potential in addressing specific challenges in this field.

To address the limitations of traditional quantum simulators and the high costs of practical quantum computers, this study proposes an innovative solution: using machine learning to simulate quantum circuits. LLMs, with their strong feature learning capabilities, can effectively process and retain dependencies in long time series data, mapping our data to quantum state vectors or density matrices. We apply this method across various quantum systems, from noise-free single qubit models to noisy two qubit models, successfully demonstrating the transition from a two qubit model to a three qubit quantum circuit. Experimental results indicate that the quantum simulator built using our method is reliable.

One major advantage of our approach is that it not only ensures results closely align with theoretical experimental outcomes but also requires lower resource consumption, exhibiting broad application potential. First, our method can output quantum state vectors or density matrices, rather than just probability values. Second, models constructed using our method can expand from low-dimensional to high-dimensional spaces, with outcomes that closely approach theoretical values. These advantages highlight the effectiveness and versatility of our proposed method, making it applicable to various quantum systems and advancing research and development in the field of quantum computing.

The structure of this paper is as follows: Section \ref{sec:llm_framework} provides a brief overview of the LLM framework used in this study and details the training process for simulating quantum circuits with LLMs. Section \ref{sec:results} presents the numerical results and analysis, focusing on the single-qubit quantum simulator model (Section \ref{sec:one_qubit}) and the two-qubit quantum simulator model (Section \ref{sec:tow_qubit}). Section \ref{sec:three_qubit} discusses the extension of the simulator, demonstrating how to construct a three-qubit quantum circuit using the two-qubit quantum simulator and presenting experimental results that test this extension method. Section \ref{sec:noise_conditions} addresses the application of the quantum simulator under realistic noise conditions, investigating its performance through experiments conducted on actual quantum devices. Finally, Section \ref{sec:discussion} summarizes the limitations of this study and discusses potential directions for future research.

\section{Methodology}
\label{sec:llm_framework}

In this study, we propose and validate a framework based on large language models (LLMs) for modeling and predicting the complex relationships between quantum circuit parameters and their corresponding quantum states. As advanced natural language processing tools, LLMs are widely used in fields such as natural language generation, translation, and dialogue due to their powerful capabilities in processing sequential data. By treating the rotation gate parameters (e.g., angles) of quantum circuits as input sequences and utilizing the self-attention mechanism of LLMs to model their interdependencies, we successfully applied this framework to the prediction of quantum states.

\begin{figure*}[!htp]
    \centering
    \includegraphics[scale=0.25]{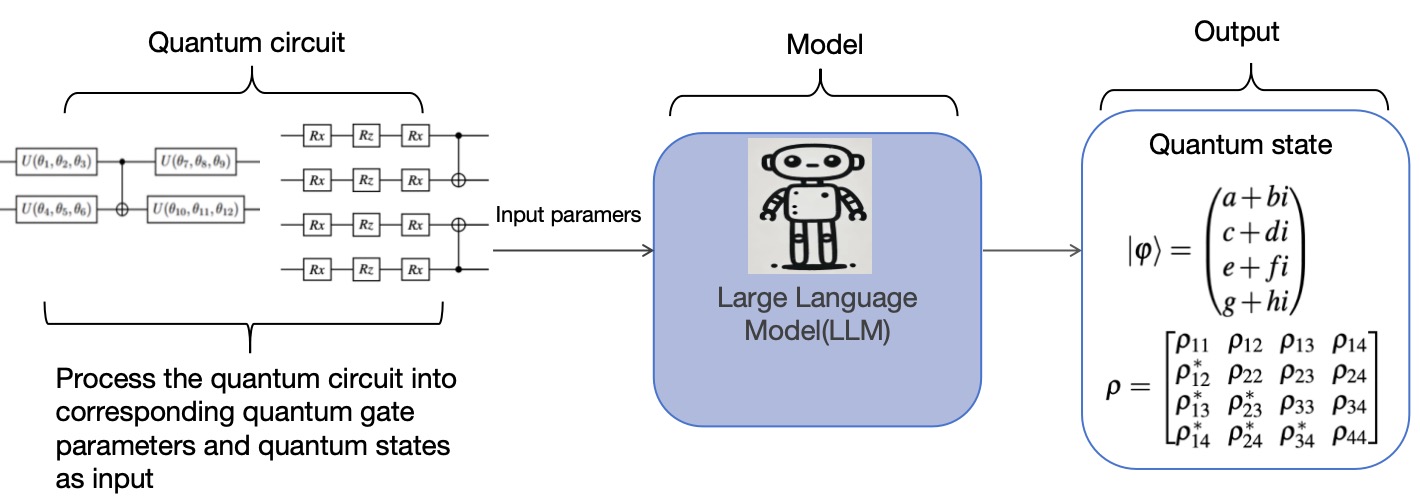}
    \caption{\label{llm_training_process}The flowchart of the model training process. First, the quantum circuit is constructed, and the corresponding quantum gate parameters and quantum states are extracted as training data. These parameters are then input into the LLM for training, and the model ultimately outputs the corresponding quantum states. Our model serves as an efficient and adaptive intermediary between the quantum circuit and quantum states.}
\end{figure*}

Specifically, the input to the model consists of the rotation angles of each quantum gate, while the output from the LLM includes the predicted quantum state, represented by the quantum state vector and the distribution of the density matrix. As illustrated in Figure \ref{llm_training_process}, this figure visually depicts the complete process from the input of quantum circuit parameters to the LLM's output of the corresponding quantum state. The core of this process lies in the LLM's ability to efficiently capture the complex interdependencies of quantum circuit parameters through its multi-layer self-attention mechanism, thereby achieving accurate modeling of quantum states in a high-dimensional space.

Compared to traditional machine learning methods, such as neural networks or classical regression models, LLMs exhibit stronger adaptability and predictive capability in quantum circuit simulations. While neural networks can handle certain nonlinear relationships, they often face limitations in generalization or computational efficiency when dealing with high-dimensional, complex quantum systems. Classical regression models, on the other hand, are typically used for linear problems and struggle to capture the nonlinear and complex interactions within quantum circuits. In contrast, LLMs can globally capture the interactions between quantum gates through their self-attention mechanism, and with their deep, stacked structure, they demonstrate superior predictive and generalization capabilities.

\begin{figure*}[!htp]
    \centering
    \includegraphics[scale=0.25]{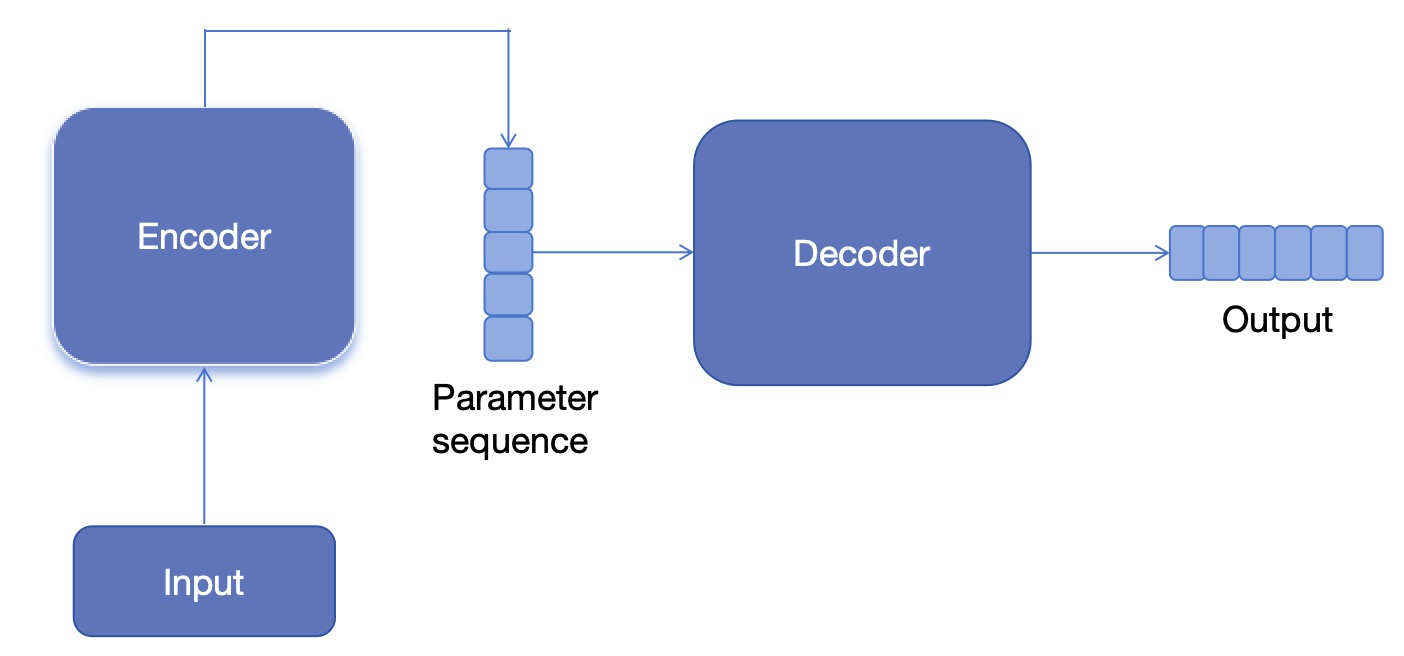}
    \caption{\label{llm_model_architecture}Our LLM model architecture consists of an encoder and a decoder. The encoder transforms the input into a sequence and passes it to the decoder. The decoder uses a self-attention mechanism to evaluate the correlations between the two inputs, generating a learned distribution. The final output is obtained by sampling from this distribution.}
\end{figure*}

As shown in Figure \ref{llm_model_architecture}, the proposed model structure includes two components: an encoder and a decoder \cite{llm7,zhuc1,llm8}. The encoder's role is to convert the input quantum circuit parameters into concise fixed-length representations, capturing the latent relationships within the input data. The decoder is used to generate the output target sequence, which can be a quantum state vector or a density matrix. The collaboration between the encoder and decoder allows the model to extract key information from the input features and effectively utilize this information when generating target parameters, thereby improving the accuracy and reliability of the generated quantum states.

Our model employs a self-attention mechanism \cite{attention1}, which dynamically adjusts weights to focus on the most relevant parts of the input data needed for the current prediction. Through the self-attention mechanism, the model can efficiently process the input sequence of quantum circuit parameters and generate the corresponding target quantum states. In the specific implementation, the continuous parameters of the quantum circuit (such as rotation angles) are treated as an input sequence, with the model using a multi-layer neural network to map these parameters into learnable representations, ultimately producing accurate predictions of the target quantum states.

To train this framework effectively, we rely on the reasonable design of the quantum circuit structure and the preparation of high-quality training data. This training data not only includes combinations of various quantum circuits but also covers the corresponding target quantum states or density matrices, which are crucial for the model to learn the complexities of quantum state distributions. During the training process, we employed a technique known as "teacher forcing" to accelerate convergence and improve model stability. This technique is commonly used in sequence generation tasks, where actual labels, rather than the model's predictions from the previous time step, are used as input during training, effectively enhancing the training speed and accuracy of the model.

Additionally, to enhance the model's autoregressive capabilities in quantum state simulation tasks, we introduced an autoregressive mechanism in the decoder. Autoregression is a time series model that uses previous outputs as inputs to iteratively predict the next value in the sequence. In our model, this mechanism not only improves the model's ability to generate continuous quantum states but also enhances its performance when handling complex quantum systems.

The objective of the model is to minimize the mean squared error (MSE), which serves as a loss function measuring the difference between the model's predicted values and the actual quantum states. The formula for MSE is as follows:

\[
\text{MSE} = \frac{1}{n} \sum_{i=1}^{n} (y_i - \hat{y}_i)^2
\]

where \(y_i\) represents the actual value, \(\hat{y}_i\)represents the predicted value, and 
\(n\) is the number of samples. By minimizing the mean squared error, the model's prediction accuracy and generalization capability are improved, ensuring that it can provide high-quality quantum state predictions for previously unseen quantum circuits.

\section{Results}~{}
\label{sec:results}

In this section, we will present our main technical contributions. Our contribution lies in the application of machine learning techniques to simulate quantum circuits. Based on this approach, we have developed both noise-free quantum simulators and noise-inclusive quantum simulators.

\subsection{1-Qubit Quantum Simulator Model}~{}
\label{sec:one_qubit}

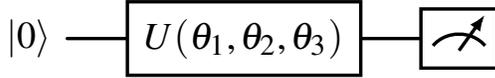
\begin{figure}[H]
	\centering
 \scalebox{1.6}{
\begin{quantikz}
  \lstick{$\ket{0}$} & \gate{U(\theta_{1}, \theta_{2}, \theta_{3})} &\meter{}\\
\end{quantikz}}
\caption{\label{u_gate} 1-qubit arbitrary quantum circuit. The quantum circuit contains a single-qubit rotation gate $U(\theta_{1}, \theta_{2}, \theta_{3})$ and measurement.}
\end{figure}

To construct a 1-qubit quantum simulator, we first need to obtain training data. To achieve this, we designed a large number of random single-qubit quantum circuits, which include U gates with random parameters, as shown in Figure \ref{u_gate}. Each $U(\theta_1, \theta_2, \theta_3)$ gate contains three parameters, $\theta_1, \theta_2$ and $\theta_3$, corresponding to the rotation angles around three axes of the Bloch sphere \cite{quantum1,quantum2}. The random combinations of these parameters can reach any point on the Bloch sphere, representing any quantum state.The mathematical expression of U is \cite{quantum1}:
\begin{equation}
U(\theta_1, \theta_2, \theta_3)=
\begin{pmatrix}
   e^{i(\frac{\theta_1}{2}-\frac{\theta_2}{2})}\cos \frac{\theta_3}{2} & 
   -e^{i(\frac{\theta_1}{2}+\frac{\theta_2}{2})}\sin \frac{\theta_3}{2} \\
   e^{i(\frac{\theta_1}{2}-\frac{\theta_2}{2})}\sin \frac{\theta_3}{2} & e^{i(\frac{\theta_1}{2}+\frac{\theta_2}{2})}\cos \frac{\theta_3}{2}
 \end{pmatrix}.
 \end{equation}
We run these quantum circuits on the quantum simulator and record the final quantum state vector of each circuit. For example, applying the $U(\theta_1, \theta_2, \theta_3)$ gate to the initial state $|0\rangle$,
\begin{equation}
U(\theta_1, \theta_2, \theta_3)
\begin{pmatrix}
   1\\
   0
 \end{pmatrix}=\begin{pmatrix}
   a+bi \\
   c+di
 \end{pmatrix},
 \end{equation}
this process yields a large number of arbitrary two-dimensional vectors, each containing four parameters: real parts $(a, c)$ and imaginary parts $(b, d)$. For the noise-free 1-qubit case, we directly record the rotation angle parameters $\theta_1, \theta_2$ and $\theta_3$ associated with the quantum gate operations in each circuit, as well as the resulting vector parameters $(a, b, c, d)$, to form our training dataset.

To encompass arbitrary quantum states, we require three rotation parameters. To achieve this, we randomly generated 7,000 samples as training data. In our input data sequence, the first three parameters are the rotation angles, which serve as the model's input, while the subsequent four parameters are the real and imaginary parts of the quantum state, which serve as the model's output.

\begin{figure*}[htbp]
	\centering
	\subfigure[The trend of training accuracy over epochs.]{
		\includegraphics[scale=0.40]{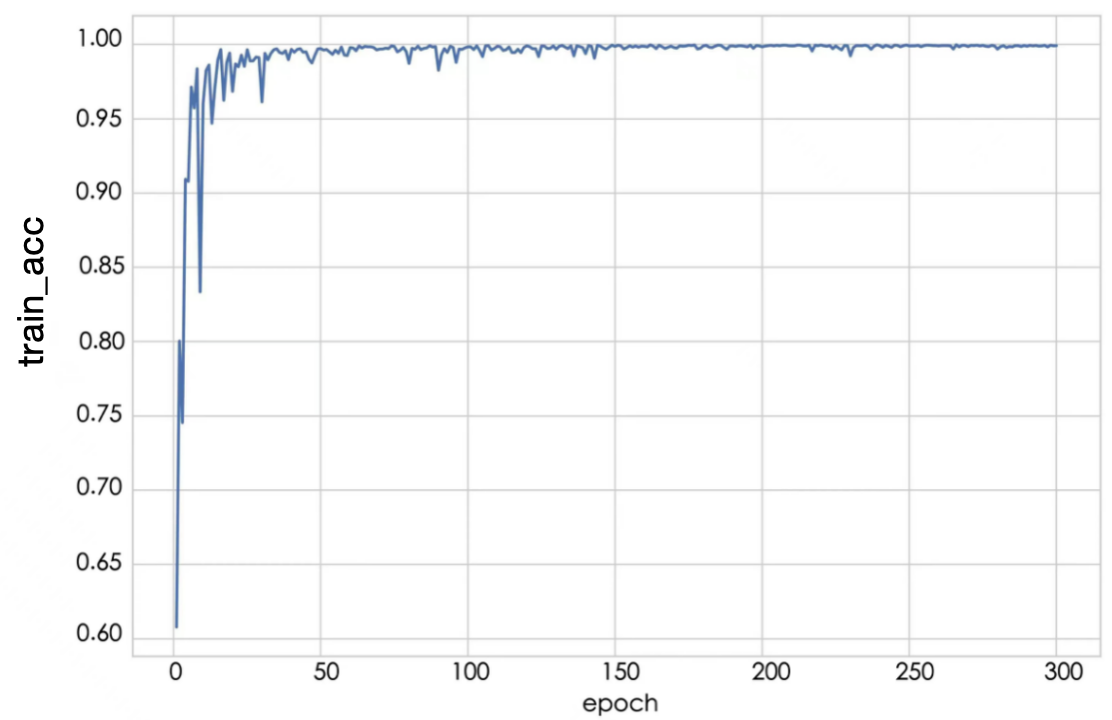}}
	\quad
	\subfigure[The variation of the loss function value over epochs.]{
		\includegraphics[scale=0.40]{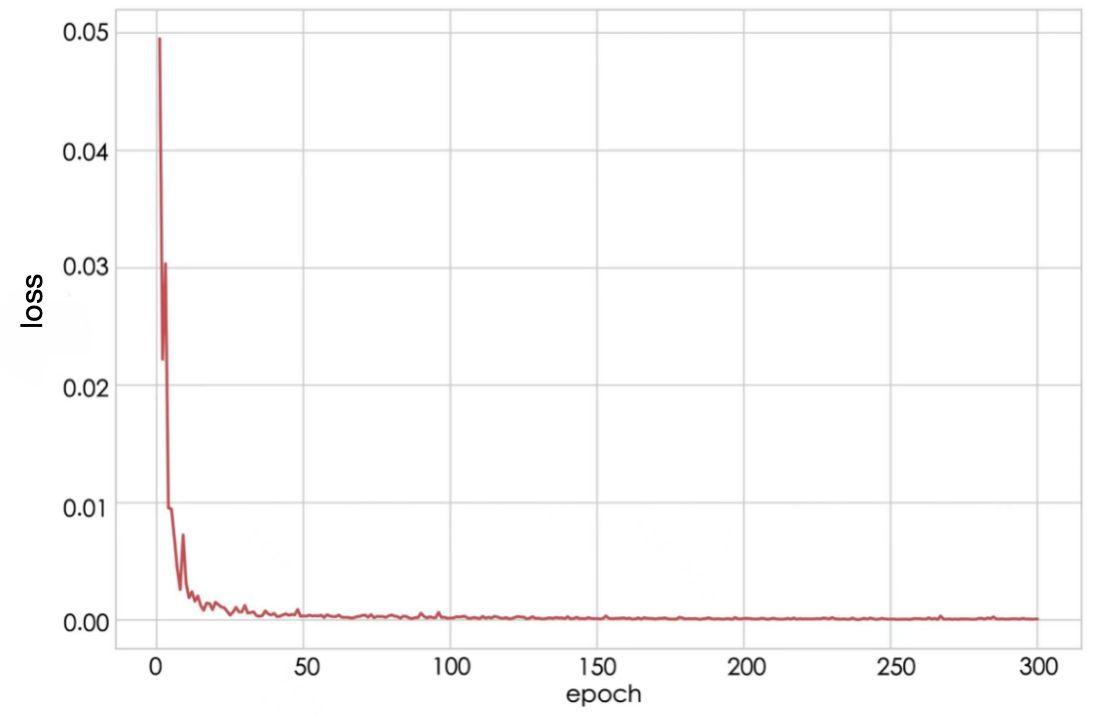}}
	\caption{Performance and Convergence of the One-Qubit Noiseless Model During Training. Figure (a) shows that as the number of training epochs increases, the model's prediction accuracy on the training data steadily improves, eventually reaching near-perfect accuracy. This indicates that the network can fit the training data well. Figure (b) shows that the loss function value gradually decreases, eventually converging to a low and stable value. This indicates that the prediction error of the network decreases continuously during the iterations, enhancing the model's generalization ability.}\label{1_qubit_no_noise_train}
\end{figure*}

We fed these samples into the LLM for training. During the training process, we performed extensive hyperparameter tuning to achieve the best results. As a result of this tuning, the model ultimately achieved an accuracy of over 99\%, as shown in Figure \ref{1_qubit_no_noise_train}.

\begin{figure}[H]
    \centering
    \includegraphics[scale=0.24]{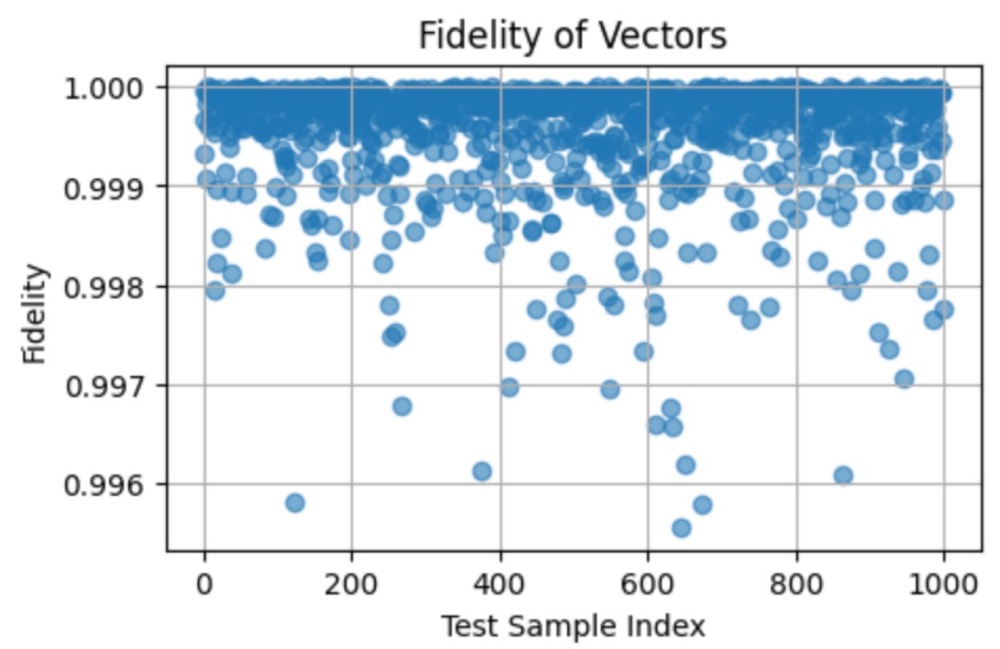}
    \caption{\label{1_qubit_no_noise}Prediction results of fidelity for different quantum state vectors by the one-qubit noiseless model on a new dataset. The comparison between the model's predictions and the true values is illustrated. The vertical axis represents the fidelity between the predicted quantum state vector and the true quantum state vector, while the horizontal axis indicates the index of each quantum state vector sample. The points are mostly clustered around 1, indicating that the predictions are largely consistent with the true values. This visually demonstrates the model's accuracy in predicting on new data.}
\end{figure}

To comprehensively evaluate the model's performance, we also tested the trained model on an independent new dataset. We constructed a new quantum state dataset with a different data distribution and used this entirely new data as input to validate the simulator's predictive capability. We used fidelity \(F=\left\vert|\langle\varphi|\varphi_{'}\rangle|\right\vert^2\) as the metric for the model's prediction accuracy, where $\varphi$ is the true quantum state vector, and $\varphi_{'}$ is the predicted quantum state vector \cite{fid1,fid2,fid3}. Figure \ref{1_qubit_no_noise} shows that the features learned from the original training data generalize well to new data. Despite being a set of entirely new quantum state vectors, the model still provides highly accurate predictions. This demonstrates the model's strong generalization ability, capable of fitting the training data and accurately predicting a broader range of unknown data.

For a 1-qubit noisy quantum simulator, the process of obtaining data is similar to the noise-free case. However, instead of recording the quantum state vectors, we now record the density matrices of the quantum states. Due to the presence of noise \cite{quantum1,quantum4}, the quantum states will be mixed states, which can be represented by density matrices \cite{quantum3}. Generally, the density matrix is defined as 
\begin{equation}
    \rho = \sum_{i}p_i|\varphi_i\rangle\langle\varphi_i|,
 \end{equation}

\begin{equation}
    p_i \geq 0, \sum_{i}p_i=1,    
 \end{equation}
which is Hermitian, meaning its conjugate transpose is equal to itself. Additionally, the trace of 
\(\rho\) (the sum of its diagonal elements) is 1\cite{quantum1}.

For a 1-qubit quantum state, the density matrix 
\(\rho\) is a 2x2 matrix, as follows:

\begin{equation}
\rho=
\begin{pmatrix}
  a_{00}& a_{01}+b_{01}i \\
   a_{01}-b_{01}i &  a_{11}
 \end{pmatrix}.
 \end{equation}
Therefore, we select \(\theta_1, \theta_2, \theta_3\) , as well as \(a_{00}, a_{01}, b_{01}, a_{11}\) as the training parameters in the 1-qubit noisy scenario. Due to the complexity of evolution in the presence of noise, we randomly generated more data and preprocessed it to train our model. The preprocessed data was then fed into the designed LLM for training.

\begin{figure}[H]
    \centering
    \includegraphics[scale=0.23]{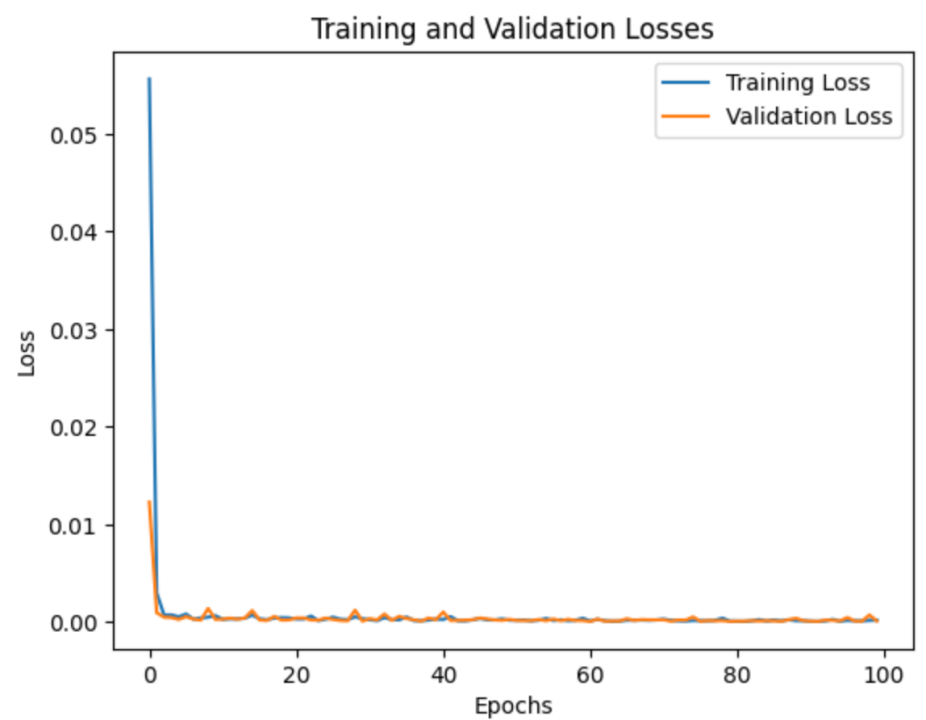}
    \caption{\label{1_qubit_noise_train}Loss value changes for the one-qubit noisy model on the training and validation sets. The blue curve represents the loss value changes on the training set, while the red curve represents the loss value changes on the validation set.}
\end{figure}

\begin{figure}[H]
    \centering
    \includegraphics[scale=0.22]{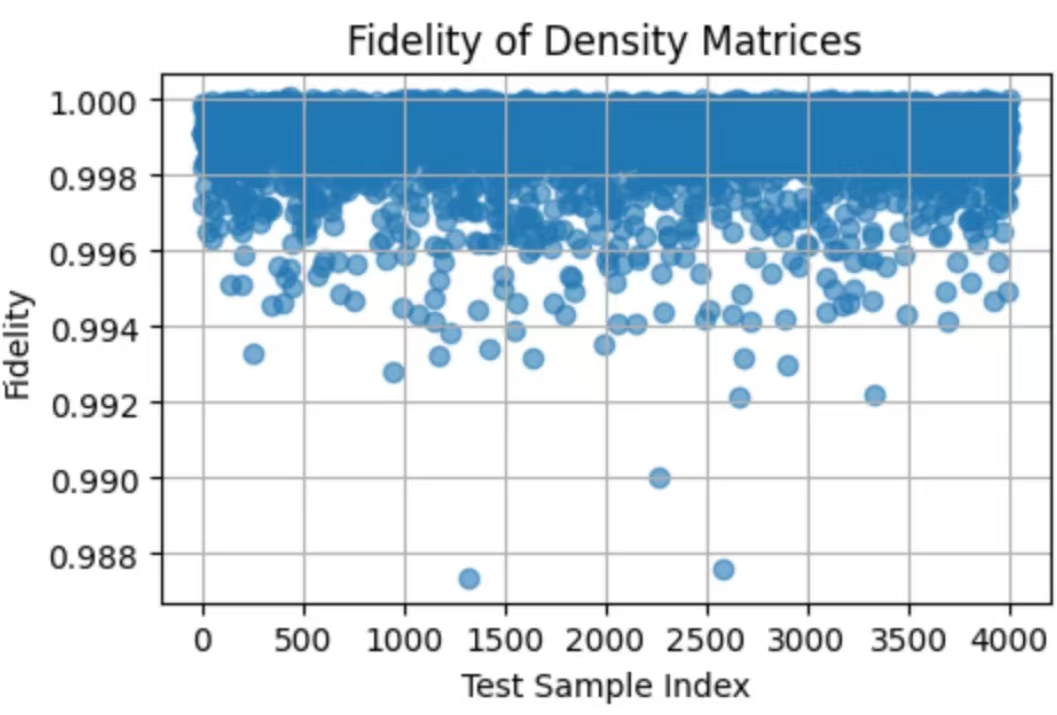}
    \caption{\label{1_qubit_noise}Prediction results of the model's fidelity for different quantum state vectors on a new dataset. The vertical axis represents quantum state fidelity, indicating how closely the predicted state matches the true state, with values closer to 1 being better. The horizontal axis represents the sample numbers. Most points are clustered above 0.987, demonstrating that the predictions are highly consistent with the true values. This clearly shows the model's accurate prediction capability on new data.}
\end{figure}

As shown in Figure \ref{1_qubit_noise_train}, with continuous hyperparameter tuning, the loss for both the training set and the validation set gradually approaches zero. Additionally, we validated the model's performance on a new dataset. We used the fidelity between two quantum states as the metric for model prediction accuracy, denoted as $F$. Its mathematical expression is\cite{fid2,quantum1}:
\begin{equation}
F(\rho,\sigma)=(tr\sqrt{\sqrt{\rho}\sigma\sqrt{\rho}})^2.
 \end{equation}
Here, $\rho$ is the density matrix of the true output, and $\sigma$ is the measured density matrix. When the simulated state $\sigma$ is sufficiently close to the true state $\rho$, this metric approaches 1. Figure \ref{1_qubit_noise} confirms that for a given distribution of unknown quantum states, the fidelity between the output and the true density matrix calculated by our model is close to 1.

\subsection{2-Qubit Quantum Simulator Model}~{}
\label{sec:tow_qubit}

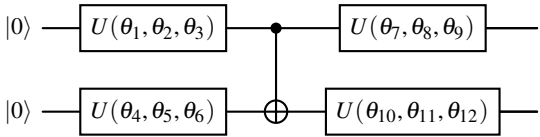
\begin{figure}[H]
    \centering
\begin{quantikz}
  \lstick{$\ket{0}$} & \gate{U(\theta_1, \theta_2, \theta_3)} &\ctrl{1}  &  \gate{U(\theta_7, \theta_8, \theta_9)}  & \qw \\
  \lstick{$\ket{0}$} & \gate{U(\theta_4, \theta_5, \theta_6)}  &\targ{}   & \gate{U(\theta_{10}, \theta_{11}, \theta_{12})}  & \qw \\
\end{quantikz}
 \caption{\label{xzx}Two-qubit special quantum circuit.}
\end{figure}

To extend the simulation to larger quantum systems, we constructed a 2-qubit quantum simulator that operates using 2 qubits. Figure \ref{xzx} illustrates the schematic for obtaining special 2-qubit quantum circuits. Each qubit undergoes independent single-qubit rotation gates $U(\theta_1, \theta_2, \theta_3)$ and $U(\theta_4, \theta_5, \theta_6)$, followed by a CNOT gate to introduce quantum entanglement. Finally, the qubits are acted upon by single-qubit gates $U(\theta_7, \theta_8, \theta_9)$ and $U(\theta_{10}, \theta_{11}, \theta_{12})$. This quantum circuit generates an arbitrary two-qubit entangled state $|\varphi\rangle$ that conforms to our circuit structure.

The quantum state vector of this entangled state is: 
\begin{equation}
|\varphi\rangle = 
\begin{pmatrix}
   a+bi \\ c+di \\
   e+fi \\ g+hi
 \end{pmatrix}.
\end{equation}

Similar to the 1-qubit case, for the noise-free scenario, our training data consists of the 12 rotation parameters 
$\theta$ for each circuit and the 8 complex parameters $(a,b,c,d,e,f,g,h)$ of the target quantum state $|\varphi\rangle$ obtained after running the quantum circuit. For the noisy scenario, the required data type is the density matrix, and the 2-qubit density matrix takes the form \cite{fid2,quantum1}:

\begin{equation}
\rho=
\begin{bmatrix}
   \rho_{11} & \rho_{12} & \rho_{13}& \rho_{14}\\
   \rho_{12}^* & \rho_{22} & \rho_{23}& \rho_{24}\\
    \rho_{13}^* & \rho_{23}^* & \rho_{33}& \rho_{34}\\
    \rho_{14}^* & \rho_{24}^* & \rho_{34}^* & \rho_{44}
 \end{bmatrix}.
 \label{density matrix1}
\end{equation}

Given the Hermitian nature of the density matrix \(\rho\), where the off-diagonal elements are complex conjugates of each other, we can express each matrix element \(\rho_{ij}\) in terms of real variables \(x_k\) and imaginary parts \(ix_k\). Specifically, we rewrite the density matrix as:

\begin{equation}
\rho=\begin{bmatrix}
   x_1 & x_2+ix_{11} & x_3+ix_{12}& x_4+ix_{13}\\
   x_2-ix_{11} & x_5 & x_6+ix_{14}& x_7+ix_{15}\\
    x_3-ix_{12} & x_6-ix_{14} & x_8& x_9+ix_{16}\\
   x_4-ix_{13} & x_7-ix_{15} & x_9-ix_{16} & x_{10}
 \end{bmatrix}.
 \label{density matrix}
\end{equation}

Therefore, we use the 12 rotation parameters $\theta$ and $x_1, x_2, x_{11}, x_3, x_{12}, x_4, x_{13}, x_5, x_6, x_{14}, x_7, x_{15}, x_8, x_9, x_{16}, x_{10}$ as the training data for the 2-qubit noisy scenario.

After obtaining the required data, the process is similar to constructing the single-qubit quantum simulator. However, due to the increased complexity of the two-qubit evolution process compared to the single-qubit case, we adjusted the LLM to accommodate the two-qubit scenario. Given that the training processes for noisy and noise-free two-qubit cases are analogous, we focused on the noisy scenario. Our training data comprises 12 rotation angles and 16 parameters of the density matrix for the final output quantum state. After preprocessing, this data was fed into the adjusted LLM for training. For convenience, we named this model “LLM-2Q Quantum Simulator.”

\begin{figure*}[!htp]
	\centering
	\subfigure[The trend of loss values with the increase in training epochs.]{
		\includegraphics[scale=0.2]{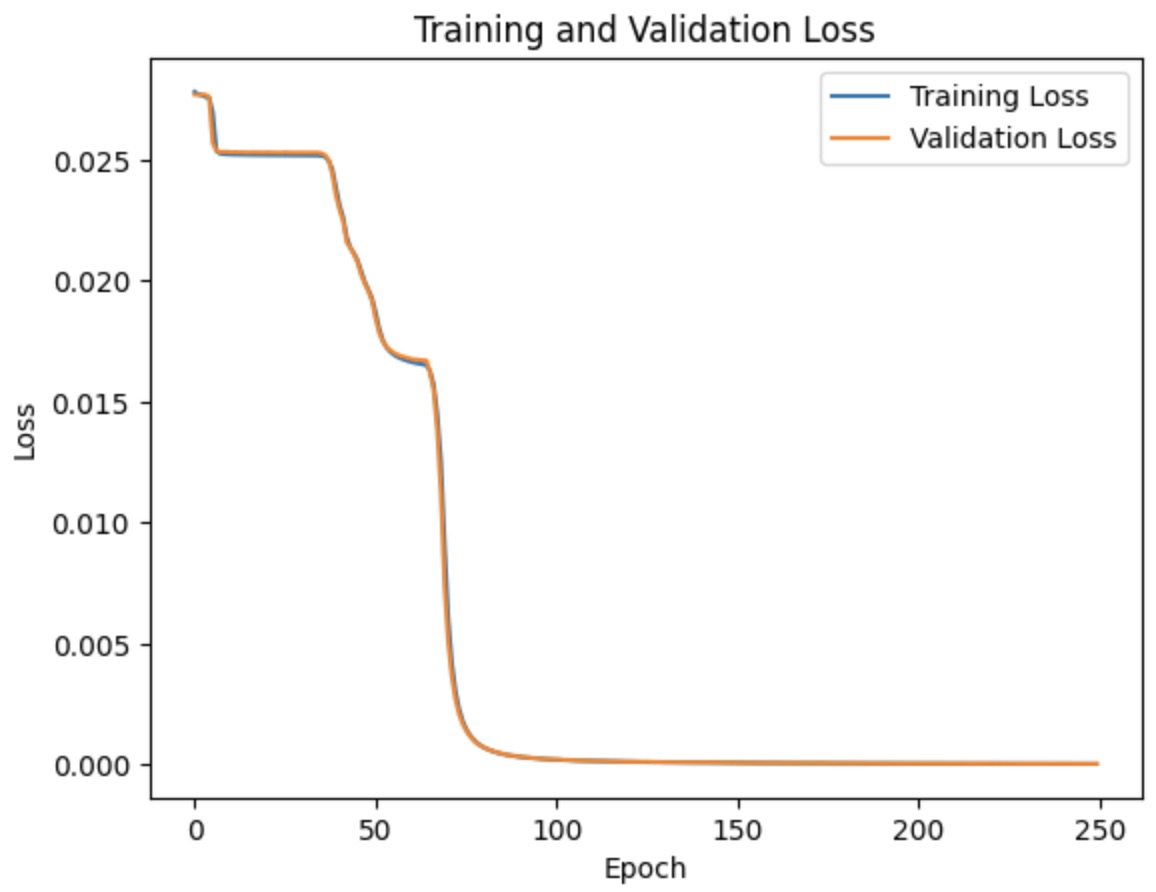}}
	\quad
	\subfigure[The model's performance on the new dataset.]{
		\includegraphics[scale=0.23]{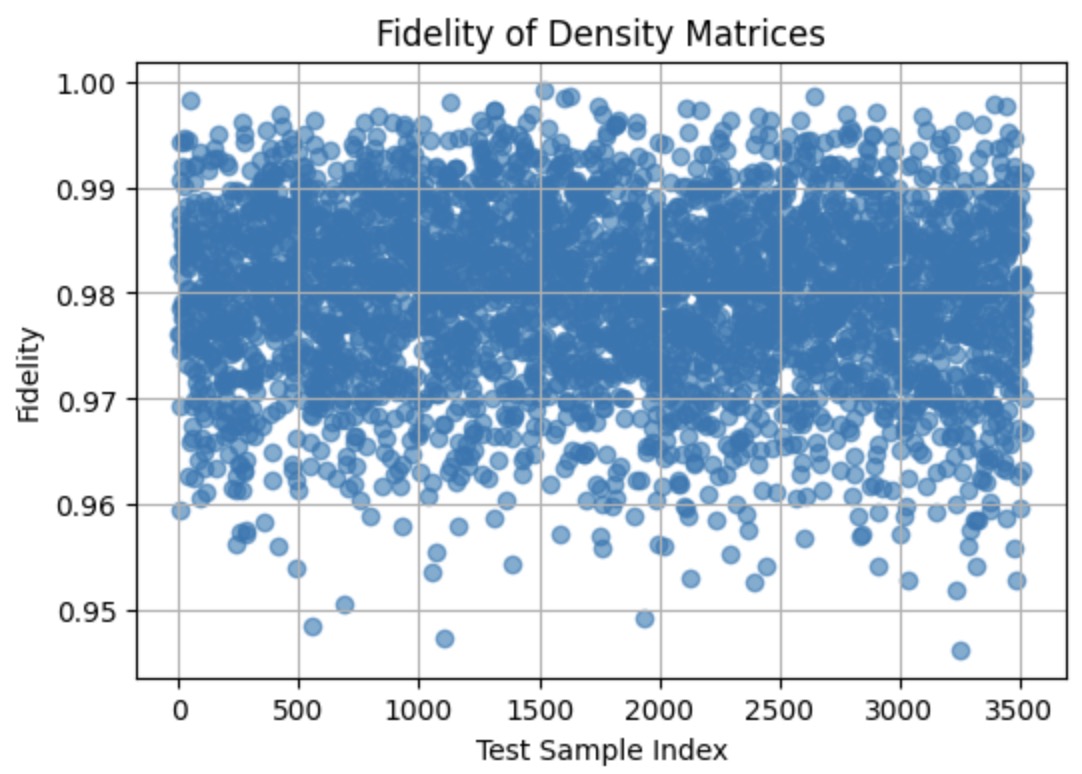}}
	\caption{Fitting performance of the two-qubit model during the training process. Figure (a) shows the loss values for the training and test sets gradually approaching zero as the number of epochs increases. Figure (b) illustrates the model's performance on new data during testing. The vertical axis represents fidelity, and the horizontal axis represents the new dataset. The minimum fidelity in the figure is above 94.7\%.}\label{2_qubit_noise}
\end{figure*}

We also validated the LLM-2Q Quantum Simulator. Figure \ref{2_qubit_noise} illustrates some results from both the training and validation processes. The figure shows that the model performs exceptionally well during training and maintains accurate predictions on new datasets. This demonstrates that the LLM-2Q Quantum Simulator exhibits strong generalization capability and accuracy in predicting complex two-qubit quantum states.

\begin{figure*}[!htp]
    \centering
    \includegraphics[scale=0.5]{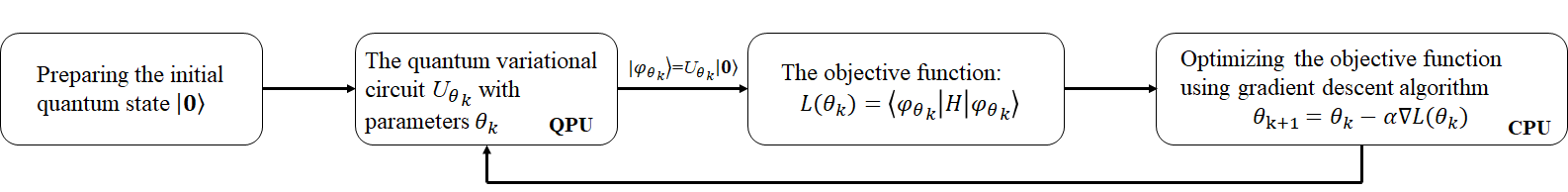}
    \caption{\label{vqe_e}Illustration of the Quantum Variational Eigensolver (VQE).}
\end{figure*}

To demonstrate the reliability of our model, we conducted a Variational Quantum Eigensolver (VQE) experiment using the LLM-2Q Quantum Simulator \cite{vqe1,vqe2}. Figure \ref{vqe_e} shows the flowchart of the VQE algorithm. The VQE algorithm primarily involves generating a parameterized quantum circuit on a quantum computer and then optimizing these parameters on a classical computer to find the optimal solution. In this experiment, we replaced the parameterized quantum circuit generated on the quantum computer with our trained model (LLM-2Q Quantum Simulator). For a 2-qubit Heisenberg model, the Hamiltonian can be expressed as \cite{hai1,hai2,hai3}:
\begin{equation}
    H = X_1\otimes X_2+Y_1\otimes Y_2+Z_1\otimes Z_2,
 \end{equation}
here, the symbol $\otimes$ represents the tensor product\cite{quantum1}, and $X_1, Y_1, Z_1$ refer to the Pauli matrices acting on the first qubit, while 
$X_2, Y_2, Z_2$ refer to the Pauli matrices acting on the second qubit.

\begin{figure}[H]
    \centering
    \scalebox{1.4}{
\begin{quantikz}
  \lstick{$q_0$} & \gate{Ry(\theta_{1})} & \ctrl{1}  &  \qw  & \qw \\
  \lstick{$q_1$} & \qw & \targ{}   & \gate{Rx(\theta_{2})}  & \qw \\
\end{quantikz}}
\caption{\label{ansatz} Ansatz of the two-qubit model, where $\theta_1$ and $\theta_2$ are parameters to be optimized.}
\end{figure}
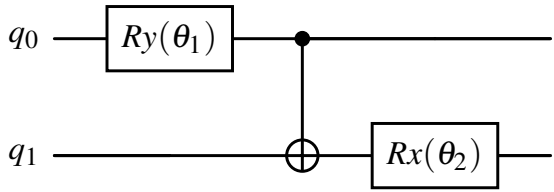

We designed a variational quantum circuit with two adjustable parameters, as shown in Figure \ref{ansatz}. In this circuit, the rotation angle of the $R_y$ gate is $\theta_1$, while the rotation angle of the $R_x$ gate is $\theta_2$. Both angles $\theta_1$ and $\theta_2$ are parameters to be optimized. By iteratively optimizing these parameters, we can determine the optimal values of $\theta_1$ and $\theta_2$ that minimize the ground state energy. In our experiments, we tested three different simulators: Qiskit's statevector\_simulator \cite{qiskit1,qiskit2}, Qiskit's AerSimulator \cite{qiskit3,qiskit4}, and our trained LLM-2Q Quantum Simulator.

\begin{figure}[H]
    \centering
    \includegraphics[scale=0.2]{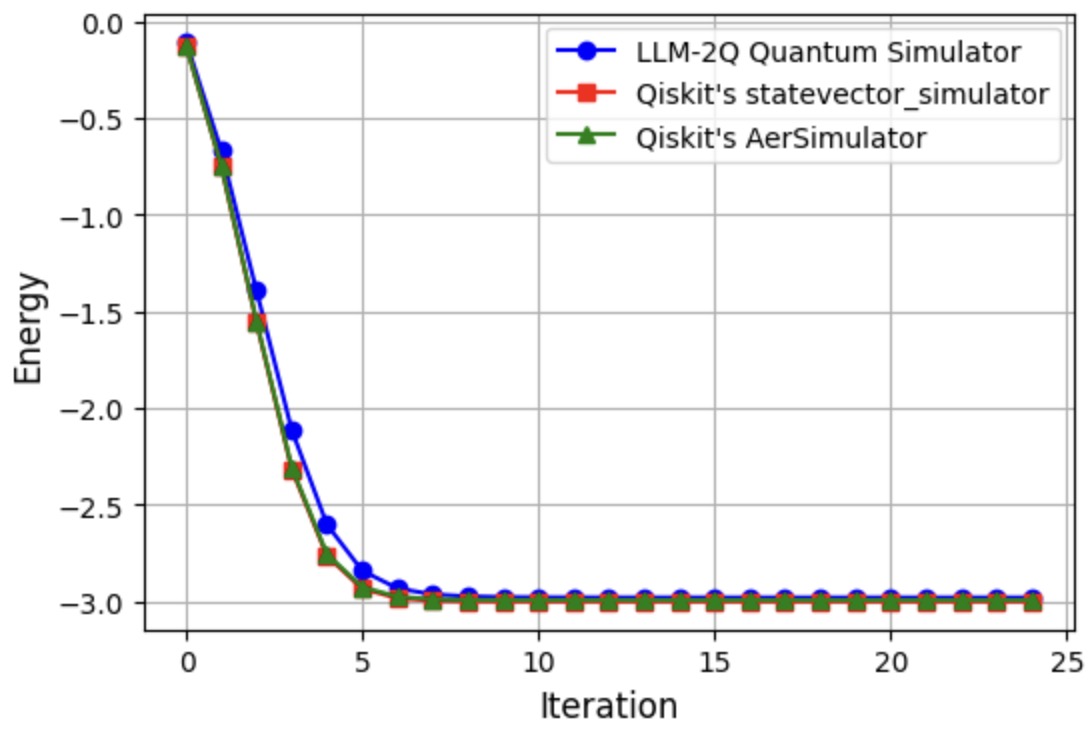}
    \caption{\label{vqe} The VQE results show the VQE energy at each iteration on different simulators. The blue dotted line represents the experimental results from our LLM-2Q Quantum Simulator. The red square line represents numerical simulations from Qiskit's statevector\_simulator. The green triangle line shows the numerical simulation results from Qiskit's AerSimulator, using the same noise parameters as our simulator.}
\end{figure}

The statevector\_simulator was used for ideal VQE experiments, providing an exact computation of the quantum state vector and allowing us to analyze the quantum circuit's performance without considering noise or errors, thus yielding results under ideal conditions. The AerSimulator is a tool in Qiskit designed for efficient simulation of quantum circuits and supports various noise models \cite{qiskit3}. In our experiment, the noise model parameters in AerSimulator were matched with those of our LLM-2Q Quantum Simulator, enabling a more intuitive assessment of our model's performance. Figure \ref{vqe} presents the experimental results. Theoretically, the ground state energy should be -3 \cite{vqe1,vqe5,vqe6}, while the result from the LLM-2Q Quantum Simulator was -2.97. These results are very close to the theoretical value, further validating the accuracy and reliability of the LLM-2Q Quantum Simulator.

Here, we introduce a model for a 2-qubit special circuit structure—the LLM-2Q Quantum Simulator. To simulate arbitrary 2-qubit quantum circuits, we extended our previous 2-qubit quantum simulator.

\begin{figure}[H]
    \centering
    \scalebox{1.2}{
\begin{quantikz}
  \lstick{} & \gate{Rx} & \gate{Rz} & \gate{Rx} &\ctrl{1}  & \qw \\
  \lstick{} & \gate{Rx} &\gate{Rz} & \gate{Rx} & \targ{}& \qw \\ 
  \lstick{} & \gate{Rx} & \gate{Rz} & \gate{Rx} & \targ{}   & \qw \\
  \lstick{} & \gate{Rx} & \gate{Rz} & \gate{Rx} & \ctrl{-1} & \qw  \\
\end{quantikz}}
\caption{\label{two_model}The two-qubit circuit model. By combining two models, any arbitrary two-qubit quantum circuit can be realized.}
\end{figure}
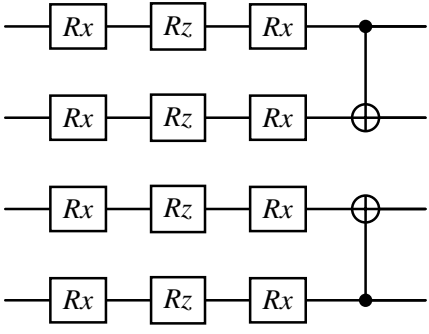

We divided the original 2-qubit circuit structure into two independent circuit structures, as shown in Figure \ref{two_model}. Each structure has 6 rotation parameters. The main difference from our previous simulator is that it could only start from the initial state $|00\rangle$ and then predict the quantum state based on the input parameters. In this extension, we trained the model to start from any initial state. Both the state parameters and the rotation parameters were used as features to train the model.

Specifically, for the noise-free case, we require 8 parameters of the initial state vector, 6 rotation angles, and 8 parameters of the resulting quantum state vector from the circuit. For the noisy case, we need the 16 parameters of the initial density matrix, the rotation angles, and the 16 parameters of the resulting density matrix from the circuit. We trained these two models with the obtained data. By combining these two models, we can simulate any 2-qubit circuit. Based on these characteristics of the model, we named the model trained in this manner the “LLM-2Q Universal Quantum Simulator”.

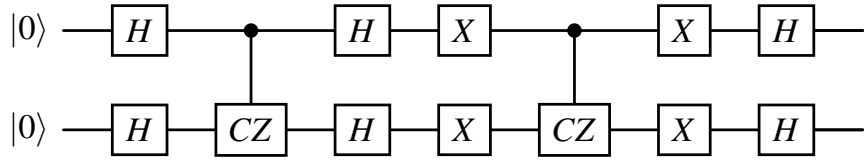
\begin{figure*}[!htp]
    \centering
    \scalebox{1.3}{
\begin{quantikz}
  \lstick{$\ket{0}$} & \gate{H} & \ctrl{1} & \gate{H}  & \gate{X}  & \ctrl{1} & \gate{X}  &\gate{H} & \qw  \\
  \lstick{$\ket{0}$} & \gate{H} & \gate{CZ} & \gate{H}  & \gate{X}  & \gate{CZ} & \gate{X}  & \gate{H}   & \qw \\  
\end{quantikz}}
\caption{\label{grover_circuit}Grover's algorithm circuit diagram..}
\end{figure*}

To validate our approach and the reliability of our models, we conducted Grover's algorithm experiments using the LLM-2Q Universal Quantum Simulator trained under both noise-free and noisy conditions and compared the results with the SpinQ Gemini mini pro quantum computer. Figure \ref{grover_circuit} shows the circuit structure of Grover's algorithm \cite{grover1,grover2,grover3}, where the Hadamard gate creates an initial uniform superposition state, and the CZ gate marks the target state we want to search for \cite{quantum1, quantum2}. Our target state is $|11\rangle$, and the CZ gate marks it by changing it to $-|11\rangle$. The operations in the circuit can be constructed using the LLM-2Q Universal Quantum Simulator, thereby implementing the entire quantum circuit.

\begin{figure*}[!htp]
    \centering
    \includegraphics[scale=0.25]{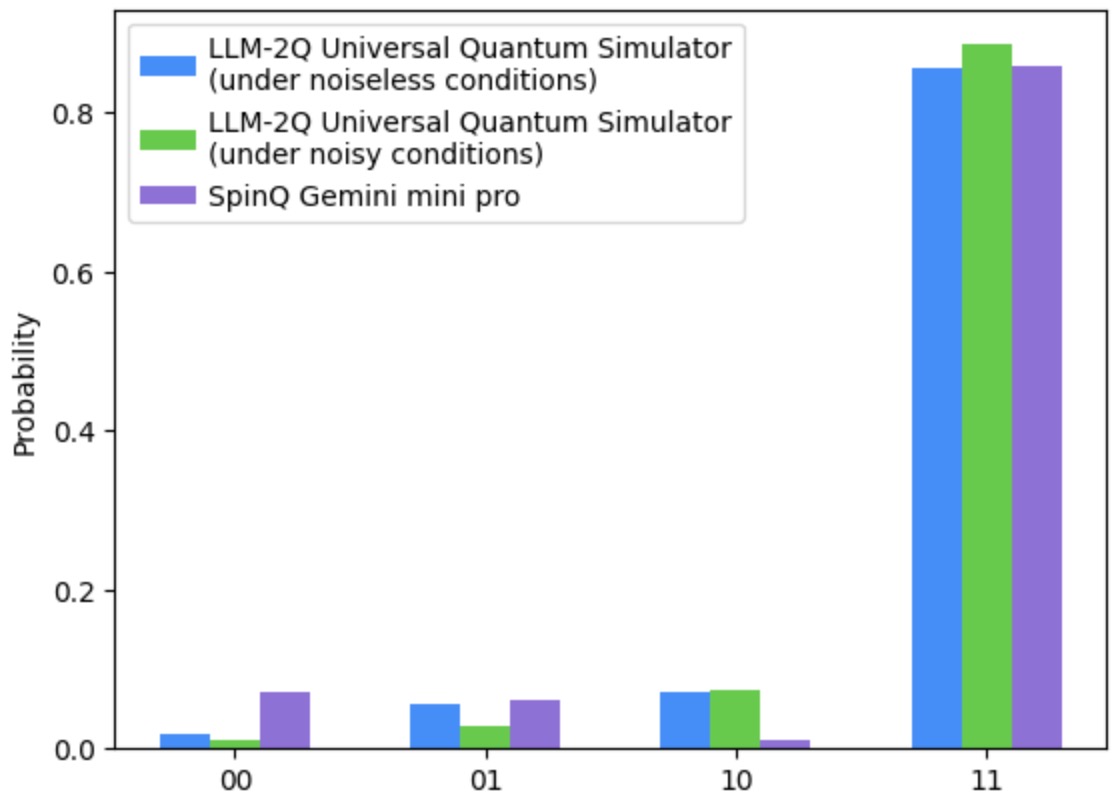}
    \caption{\label{grover} In the Grover's algorithm, the blue line represents the experimental results of the noisy LLM-2Q Universal Quantum Simulator, while the green line indicates the results of the noise-free LLM-2Q Universal Quantum Simulator. The purple line represents the experimental results of the SpinQ Gemini mini pro quantum computer.}
\end{figure*}

Figure \ref{grover} shows the results of the Grover's algorithm simulation using the LLM-2Q Universal Quantum Simulator. The blue line represents the results under noisy conditions, the green line represents the noise-free results, and the purple line indicates the experimental results of the SpinQ Gemini mini pro quantum computer. It is evident that among the three, the highest probability for the $|11\rangle$ state is achieved by the noise-free LLM-2Q Universal Quantum Simulator. Under noisy conditions, the results from the LLM-2Q Universal Quantum Simulator and the SpinQ Gemini mini pro quantum computer are similar, with both yielding a higher probability for the $|11\rangle$ state. The target state we aim to find is $|11\rangle$, indicating that our model can accurately locate the corresponding target state within an acceptable error range.

\subsection{Expansion of Quantum Simulator}~{}
\label{sec:three_qubit}

In the previous section, we introduced a method for constructing a simulator for arbitrary two-qubit circuits by combining two models. Building on this framework, we further explored how to extend this method to simulate three-qubit circuits without incurring additional training costs. By leveraging the existing two-qubit models, we can effectively construct a three-qubit simulator without the need to retrain a separate three-qubit model. This extension not only reduces training time and computational resource consumption but also significantly enhances the model's scalability, making it applicable to more complex quantum circuit simulations.

\begin{figure*}[!htp]
    \centering
    \includegraphics[scale=0.6]{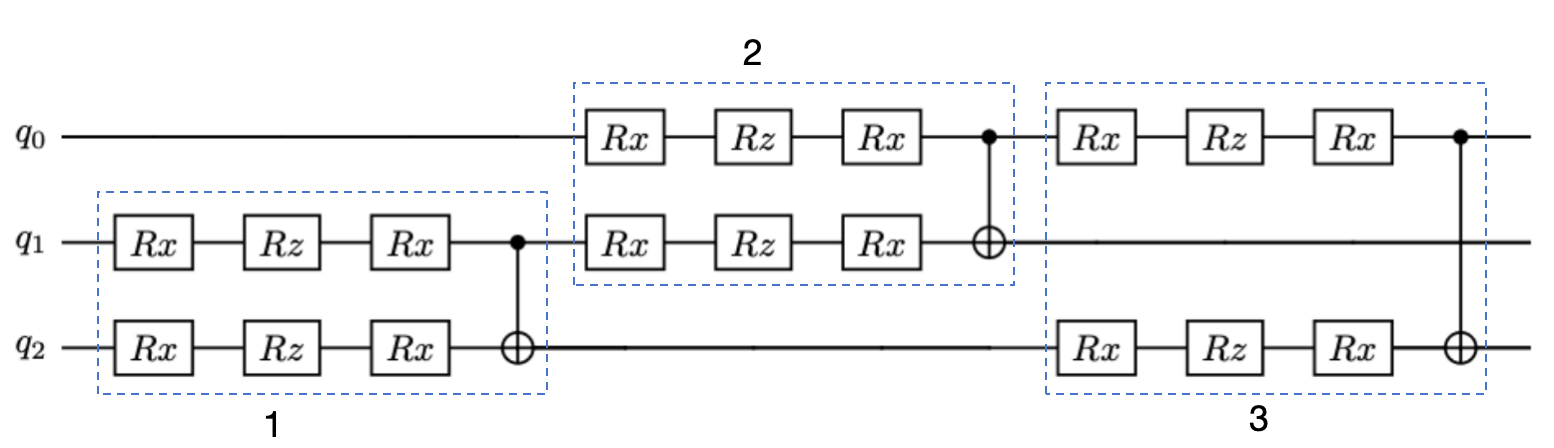}
    \caption{\label{3_qubit_circuit}Three-qubit circuit structure.}
\end{figure*}

As shown in Figure \ref{3_qubit_circuit}, we constructed a 3-qubit quantum circuit. In the figure, dashed boxes 1, 2, and 3 represent the LLM-2Q Universal Quantum Simulator. We only need to ensure that the input feature parameters meet the requirements of the LLM-2Q Universal Quantum Simulator. For the density matrix of the 3-qubit quantum circuit, we can represent it as follows \cite{quantum1}:
\begin{equation}
\rho = |0\rangle\langle0|\otimes \rho^{00}+|0\rangle \langle1|\otimes \rho^{01}+|1\rangle \langle0|\otimes \rho^{10}+|1\rangle \langle1|\otimes \rho^{11},
\label{1}
\end{equation}

where
\begin{align*}
\rho \in C^{8\times8},
\end{align*}

and
\begin{align*}
|0\rangle\langle0|,...,|1\rangle\langle1| \rightarrow q_0, \
\rho^{00}, \rho^{01}, \rho^{10}, \rho^{11} \in C^{4\times4} \rightarrow q_{12}.
\end{align*}

The matrix form of $\rho$ can be expressed as
\begin{equation}
\rho=
\begin{bmatrix}
\rho_{00} & \rho_{01} \\
\rho_{10} & \rho_{11}
\end{bmatrix} \in C^{8\times8}.
\end{equation}

We can use Equation \ref{1} to decompose the matrices $\rho_{00}$, $\rho_{01}$, $\rho_{10}$, and $\rho_{11}$ and process these four matrices into parameters suitable for the LLM-2Q Universal Quantum Simulator. These parameters are then input into the LLM-2Q Universal Quantum Simulator to obtain the output matrices. After processing the output matrices, we can recombine them to restore them to a 3-qubit density matrix. Through this iterative process of decomposition and recombination, we can simulate any 3-qubit quantum circuit. The methods for decomposition and recombination are as follows:

1. \textbf{Dashed Box 1}: $q_0\otimes q_{12}$
    
    For this case, the decomposition method satisfies Equation \ref{1}, allowing us to obtain the matrices $\rho_{00}, \rho_{01}, \rho_{10}$ and $\rho_{11}$.

2.  \textbf{Dashed Box 2}: $q_{01}\otimes q_2$

    The corresponding mathematical expression is:
\begin{equation}
\rho = \rho^{00}\otimes|0\rangle\langle 0|+\rho^{01}\otimes|0\rangle\langle 1|+\rho^{10}\otimes|1\rangle\langle 0|+\rho^{11}\otimes|1\rangle\langle 1|,
\end{equation}
where $|0\rangle\langle0|,...,|1\rangle\langle1|$ represent $q_2$, and $\rho^{01},...,\rho^{11}$ represent $q_{01}$. $\rho_{00}, \rho_{01}, \rho_{10}$ and $\rho_{11}$ are the four matrices we need to decompose.

3. \textbf{Dashed Box 3}: $q_{02}\otimes q_1$

For this case, the decomposition method satisfies:

\begin{equation}
\begin{aligned}
\rho &= (|0\rangle\langle0|)_1 \otimes \rho^{00} + (|0\rangle\langle1|)_1 \otimes \rho^{01} \\
&\quad + (|1\rangle\langle0|)_1 \otimes \rho^{10} + (|1\rangle\langle1|)_1 \otimes \rho^{11}.
\end{aligned}
\end{equation}

where the subscript 1 indicates the first qubit, and $\rho_{00}, \rho_{01}, \rho_{10}$ and $\rho_{11}$ represent the matrix composed of the zeroth qubit and the second qubit.

For the above three decomposition methods, the matrices $\rho_{00}$, $\rho_{01}$, $\rho_{10}$, and $\rho_{11}$ may not necessarily satisfy the conditions of a density matrix. To adapt them for input into the LLM-2Q Universal Quantum Simulator, we need to make the necessary adjustments. A density matrix must be Hermitian, have a trace equal to 1, and be positive semi-definite \cite{quantum1}.
\begin{itemize}
    \item $\rho_{00}$ and $\rho_{11}$ and are Hermitian.
    \item $\rho_{01}$ and  $\rho_{10}$ are non-Hermitian; however, these two matrices are Hermitian conjugates of each other. Thus, $\rho_{01}+\rho_{10}$ forms a Hermitian matrix, denoted as $S$. The difference $(\rho_{01}-\rho_{10})i$ is also a Hermitian matrix, denoted as $D$.
\end{itemize}

To ensure positive semi-definiteness, we scale the matrices $\rho_{00}, \rho_{11}$, $S$, and $D$ by a factor $\alpha$ that is less than 1. Finally, we add a unit matrix $\textit{I}$ multiplied by an appropriate coefficient $\beta$ to ensure the trace equals 1 \cite{theory1,theory2,theory3}. This results in four matrices that strictly meet the conditions of a density matrix. By inputting these four matrix parameters into the LLM-2Q Universal Quantum Simulator, the corresponding parameters are output.

To restore the matrices corresponding to $\rho_{00}, \rho_{01}, \rho_{10}$ and $\rho_{11}$ we perform the inverse operations. Specifically, we subtract the matrices added earlier (the unit matrix multiplied by the coefficient $\beta$) from each of the four matrices and then divide by the scaling factor $\alpha$ to obtain the matrices corresponding to $\rho_{00}, S, D$ and $\rho_{11}$ as $\rho_{00}^{'}, S^{'}, D^{'}$ and $\rho_{11}^{'}$, For $\rho_{01}^{'}$ and $\rho_{10}^{'}$, we use the relations:
\begin{equation}
  \rho_{01}^{'}=\frac{(S^{'}-D^{'}i)}{2}, 
\rho_{10}^{'}=\frac{(S^{'}+D^{'}i)}{2}.
\end{equation}

In this way, we acquire four $4\times4$ matrices, $\rho_{00}^{'}, \rho_{01}^{'}, \rho_{10}^{'}$ and $\rho_{11}^{'}$, representing the state of the three qubits after the corresponding gate operations. Finally, we can reconstruct the 3-qubit density matrix by substituting these four matrices back according to the earlier decomposition method. This iterative process of decomposition and recombination allows us to construct arbitrary 3-qubit quantum circuits, facilitating the extension from a 2-qubit model to the simulation of 3-qubit quantum circuits.

\begin{figure*}[!htp]
    \centering
    \includegraphics[scale=0.25]{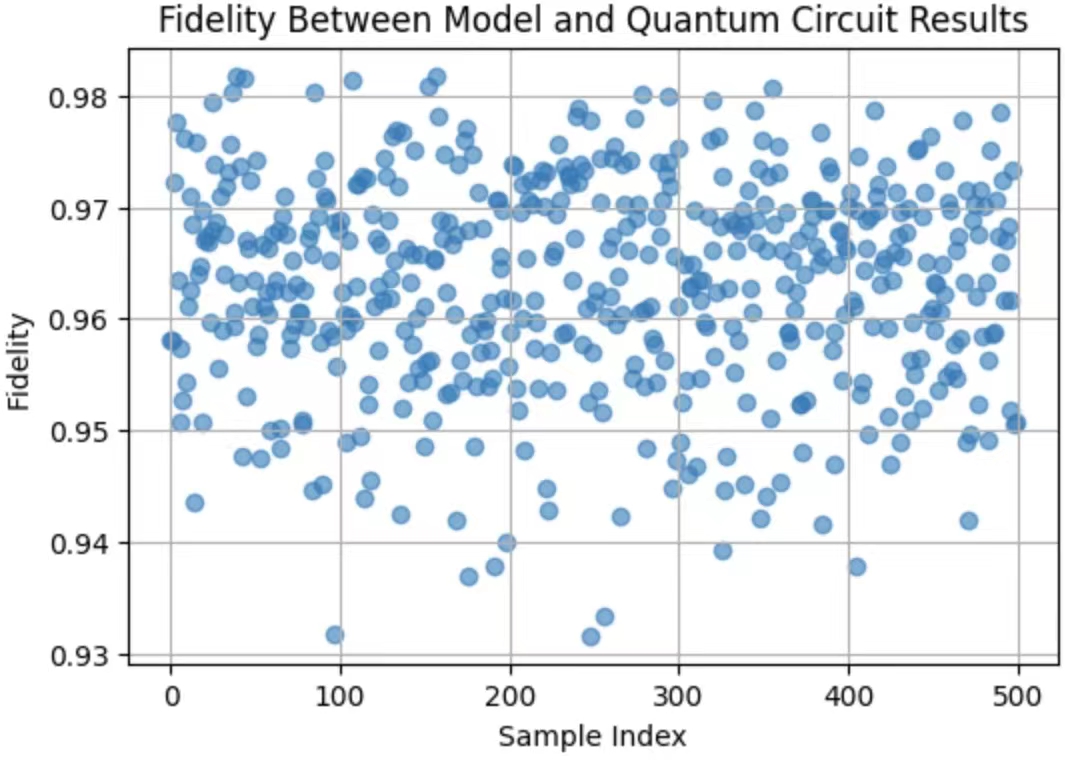}
    \caption{\label{3_qubit_fid} This figure shows the fidelity between the model results and the actual quantum circuit results for different samples. The horizontal axis "Sample Index" represents the index of each test sample, where each point corresponds to a different random 3-qubit circuit and its related quantum circuit result. The vertical axis "Fidelity Between Model and Circuit" represents the fidelity between the density matrix predicted by the model and the density matrix output by the actual quantum circuit. As can be observed from the figure, the fidelity for each test sample is above 93\%, indicating the accuracy of the model in simulating quantum states.}
\end{figure*}

To validate the performance of the model, we designed and conducted a series of experiments simulating several randomly generated three-qubit circuits. We compared the results generated by the model with the outcomes from actually running the same quantum circuits. The core metric in these experiments was the fidelity between the density matrices, which measures the similarity between the quantum states generated by the model and those produced by real hardware. Figure \ref{3_qubit_fid} presents these fidelity results, where the horizontal axis represents different random three-qubit circuit samples, and the vertical axis indicates the fidelity between the model-generated density matrices and the actual results. The experimental results show that the model exhibits high fidelity across all test samples, with fidelities consistently exceeding 93\%, demonstrating the model's high accuracy and stability in quantum state simulation. Regardless of the complexity of the circuit structure, the model exhibited highly consistent performance, further confirming its adaptability to different quantum state evolution pathways.

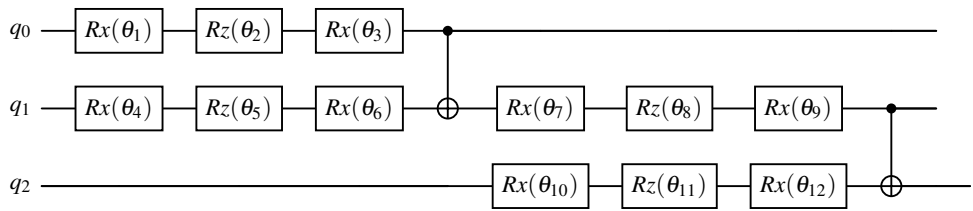
\begin{figure*}[!htp]
     \centering
    \scalebox{0.9}{
 \begin{quantikz}
 \lstick{$q_0$} & \gate{Rx(\theta_{1})} & \gate{Rz(\theta_{2})} & \gate{Rx(\theta_{3})} & \ctrl{1}  & \qw & \qw & \qw & \qw &  \qw   \\
   \lstick{$q_1$} & \gate{Rx(\theta_{4})} & \gate{Rz(\theta_{5})} & \gate{Rx(\theta_{6})} & \targ{} & \gate{Rx(\theta_{7})} & \gate{Rz(\theta_{8})} & \gate{Rx(\theta_{9})} &\ctrl{1} & \qw \\
   \lstick{$q_2$} & \qw &\qw & \qw & \qw & \gate{Rx(\theta_{10})} & \gate{Rz(\theta_{11})} & \gate{Rx(\theta_{12})} & \targ{} & \qw & \qw\\
 \end{quantikz}}
 \caption{\label{3_qubit_vqe} Parameterized circuit for a three-qubit Variational Quantum Eigensolver (VQE). Multiple rotation gates $Rx(\theta_i)$ and $Rz(\theta_j)$ are applied to each qubit line, with adjustable parameters $\theta$ to optimize the quantum state. The circuit also includes controlled-NOT (CNOT) gates to introduce entanglement between qubits. This structure allows the circuit to minimize the expectation value of the Hamiltonian by tuning the parameters, thereby approximating the ground state energy of the system.}
\end{figure*}

\begin{figure*}[!htp]
    \centering
    \includegraphics[scale=0.5]{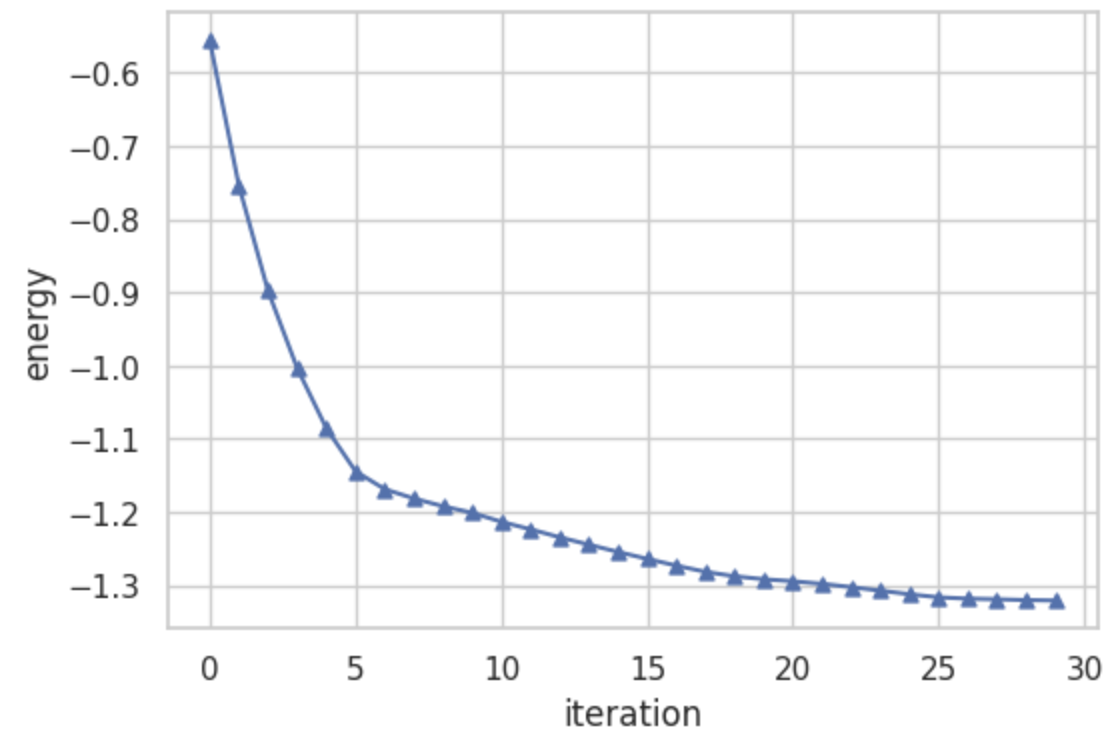}
    \caption{\label{3_qubit_vqe_result} The figure illustrates the energy convergence process in the Variational Quantum Eigensolver (VQE) experiment. The horizontal axis represents the number of optimization iterations, while the vertical axis shows the estimated ground state energy of the system after each iteration. As the number of iterations increases, the parameters are gradually optimized, and the energy approaches the ground state energy of the target Hamiltonian.}
\end{figure*}

To further assess the capabilities of this extended model, we conducted an experiment based on the Variational Quantum Eigensolver (VQE) to evaluate the model's performance in solving the ground-state energy problem of complex quantum systems. The VQE experiment aims to approximate the ground-state energy of a target Hamiltonian by optimizing a parameterized quantum circuit. The VQE circuit, as shown in Figure \ref{3_qubit_vqe}, includes several rotation gates $Rx(\theta_i)$ and $Rz(\theta_j)$ and introduces quantum entanglement via controlled-NOT (CNOT) gates, ensuring that the circuit captures the complex quantum state evolution. The target Hamiltonian used in the experiment is:

\begin{equation}
    H = X_1\otimes X_2\otimes X_3 +Y_1\otimes Y_2\otimes Y_3+Z_1\otimes Z_2\otimes Z_3,
 \end{equation}

where $X_i$, $Y_i$, and $Z_i$ are Pauli operators acting on the $i$-th qubit. Theoretically, the ground-state energy of this Hamiltonian is $-1.4$. As shown in Figure \ref{3_qubit_vqe_result}, after VQE optimization, the model estimated the ground-state energy to be $-1.33$. Although there is a deviation from the theoretical value, the result still demonstrates the model’s ability to approximate complex quantum systems, proving its effectiveness in variational quantum algorithms.

Overall, the extended three-qubit simulator, based on the two-qubit model, exhibits superior performance in multiple aspects. First, by extending the two-qubit architecture, the model efficiently simulates three-qubit circuits without retraining, significantly improving simulation efficiency. Second, the experimental results show that the model achieves high accuracy and stability in both quantum state simulation and variational quantum eigensolver tasks, further validating its potential in simulating more complex quantum systems. This lays a solid foundation for future applications of large language models in simulating complex quantum circuits and provides new directions for research in quantum computing.

\subsection{The model's simulation under realistic noise conditions}~{}
\label{sec:noise_conditions}

In the preceding sections, we introduced a single-qubit quantum simulator that incorporates noise \cite{quantum1,quantum4,noise}, specifically relaxation and dephasing noise. While this provides a useful framework for understanding quantum noise, it still represents an idealized scenario that does not fully capture the complexities of a real quantum computer. To address this limitation, this section presents an experimental approach to obtaining data under realistic noise conditions by running experiments on an actual quantum device.

\begin{figure*}[!htp]
    \centering
    \includegraphics[scale=0.30]{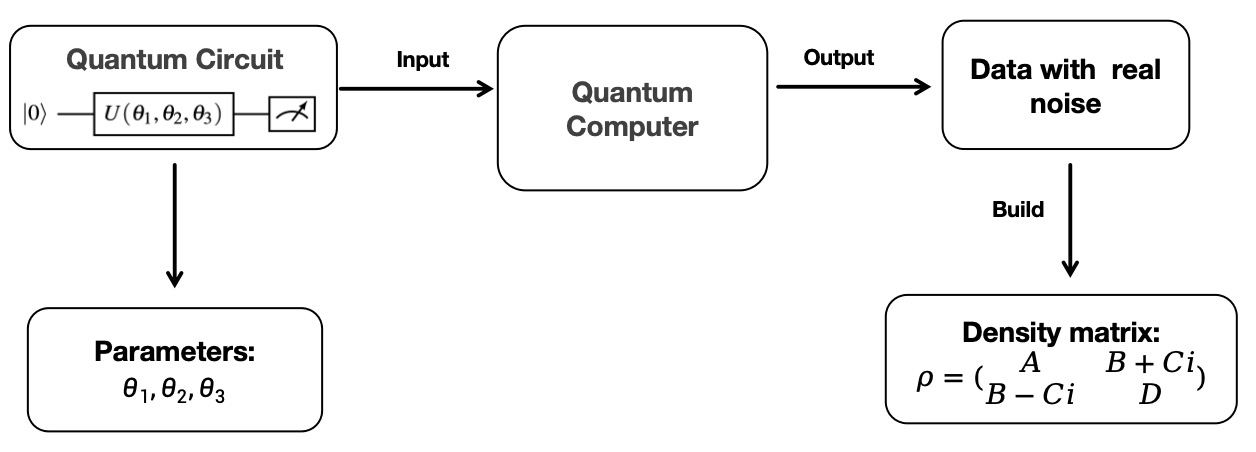}
    \caption{\label{Flowchart}Flowchart for acquiring real quantum computer data.}
\end{figure*}

As depicted in Figure \ref{Flowchart}, we first construct a single-qubit quantum circuit and input it into a quantum computer. The quantum computer then outputs the corresponding density matrix for that circuit. By generating a large number of random circuits and inputting them into the quantum computer, we collect the resulting outputs, which serve as the training data for our model. It is important to note that a real quantum computer can only output probabilities (i.e., the diagonal elements of the density matrix) rather than the complete density matrix. To reconstruct the density matrix, we employ state tomography and convex optimization techniques, where we reconstruct the matrix from the probabilities output by the quantum computer and then apply mathematical operations to optimize it to meet the criteria for a valid density matrix. A more detailed explanation of these methods is provided in the appendix A.

\begin{figure*}[!htp]
	\centering
	\subfigure[The trend of loss values with the increase in training epochs.]{
		\includegraphics[scale=0.25]{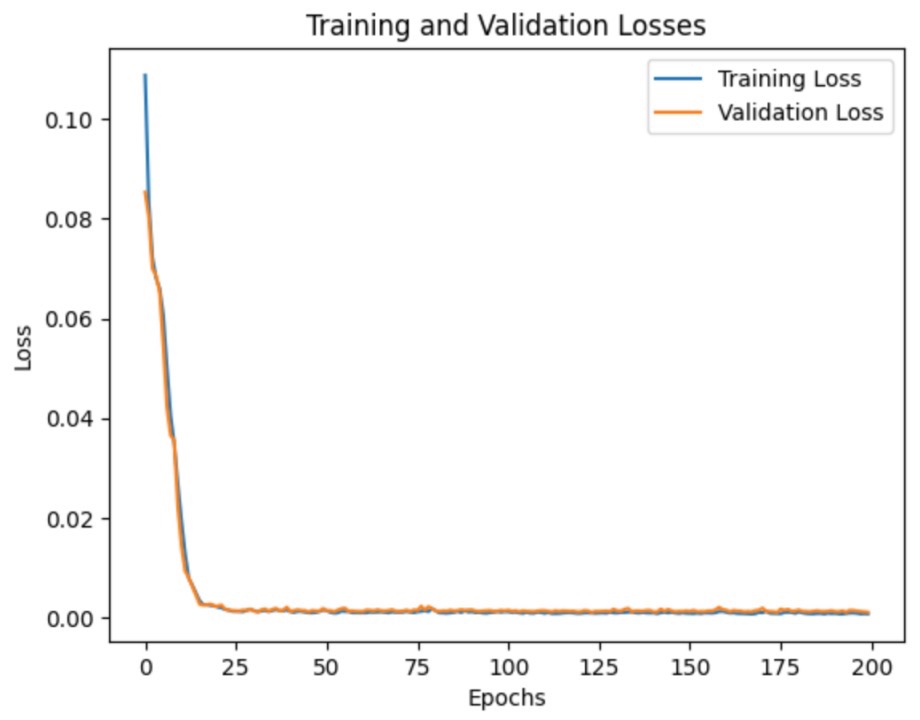}}
	\quad
	\subfigure[The model's performance on the new dataset.]{
		\includegraphics[scale=0.25]{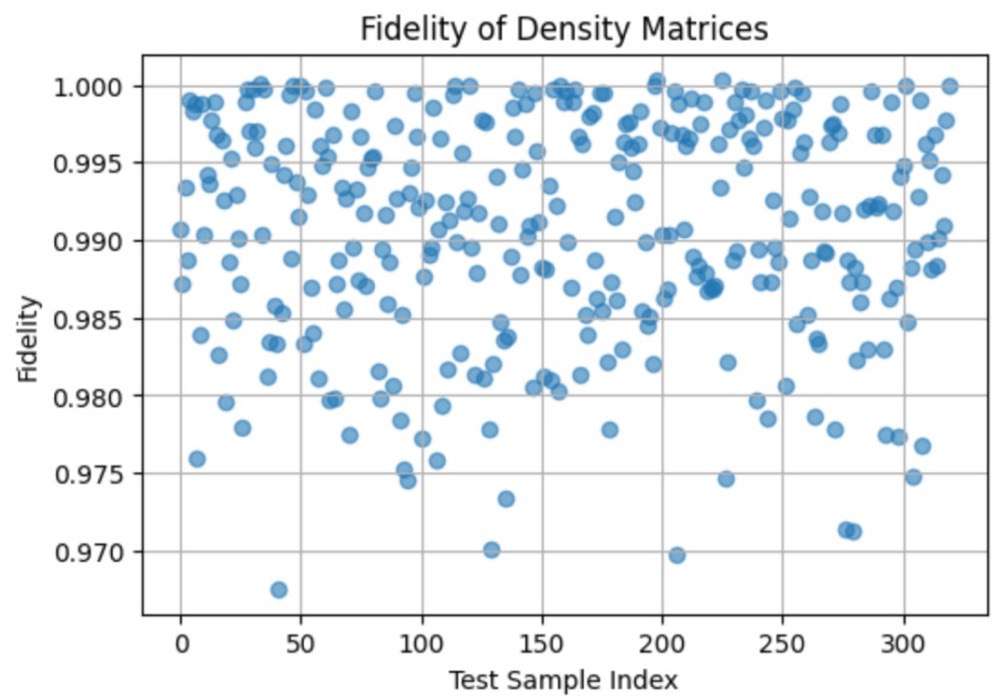}}
	\caption{Fitting performance of the two-qubit model during the training process. Figure (a) shows the loss values for the training and test sets gradually approaching zero as the number of epochs increases. Figure (b) illustrates the model's performance on new data during testing. The vertical axis represents fidelity, and the horizontal axis represents the new dataset. The minimum fidelity in the figure is above 96.7\%.}\label{1_qubit_true_noise}
\end{figure*}

Given the complexity of quantum evolution in noisy environments, we generated additional data to train our model and applied preprocessing steps before inputting the data into the designed LLMs for training. The results of the training and testing processes are shown in Figure \ref{1_qubit_true_noise}. As observed, the model performs well even in the presence of highly complex noise, demonstrating its robustness and reliability when applied to new datasets. This suggests that our model is effective in simulating quantum circuits under realistic noisy conditions.

\section{Discussion}~{}
\label{sec:discussion}

This study integrates large language models (LLMs) to develop a novel quantum state simulator, which was extensively tested using Grover's algorithm and the Variational Quantum Eigensolver (VQE). The test results show that the simulator exhibits high reliability when handling quantum circuits in both noise-free and noisy environments, aligning well with theoretical expectations.

Despite these significant achievements, the simulator still has some limitations. Firstly, although it performed well in three-qubit experiments, when scaling to larger numbers of qubits, the model may encounter resource bottlenecks, and efficiency optimization is required. Secondly, while the model maintains high accuracy under current noise conditions, its precision may be affected when dealing with more complex noise models. Therefore, future research should focus on enhancing the model's robustness to address a wider variety of quantum noise environments.

The study also demonstrates that the LLM-based simulator can efficiently reproduce both quantum state vectors and density matrices, enabling comprehensive simulations of complex quantum systems. This capability is particularly valuable for quantum computing experimental simulations, as it significantly reduces resource consumption during the initial stages of experiments. By performing efficient simulations in advance, researchers can better assess the performance of quantum circuits in practical scenarios, optimize experimental design, and minimize trial-and-error costs.

The introduction of LLMs offers new research perspectives for the quantum computing field, particularly in resource-constrained situations. This study highlights the tremendous potential of LLMs in quantum state simulation, especially in terms of model scalability. Additionally, LLMs provide researchers with an alternative pre-experimental testing method, lowering the cost barriers of expensive quantum computing experiments. As the model continues to expand, it is expected to more easily predict the complexity of actual quantum computing experiments. More importantly, the high computational speed of LLMs opens new possibilities for improving and optimizing quantum algorithms, particularly those involving large numbers of quantum gate operations.

Future research could focus on several areas: first, optimizing the structure of LLMs to ensure their continued efficiency when simulating larger numbers of qubits; second, further improving the simulator's robustness in complex quantum noise environments, particularly by fine-tuning the model for more detailed simulations of noise conditions in real quantum devices. Furthermore, given the flexibility of LLMs, future studies could explore their application in a broader range of quantum algorithms, such as simulating quantum many-body systems and quantum machine learning. These research directions would not only enhance the multifunctionality of quantum simulators but also drive widespread applications in both theoretical and practical aspects of quantum computing.

\section{Acknowledgement}
This work is supported by National Natural Science Foundation of China under Grant No. 12105195.

\bibliography{sample}

\begin{thebibliography}{53}%
\makeatletter
\providecommand \@ifxundefined [1]{%
 \@ifx{#1\undefined}
}%
\providecommand \@ifnum [1]{%
 \ifnum #1\expandafter \@firstoftwo
 \else \expandafter \@secondoftwo
 \fi
}%
\providecommand \@ifx [1]{%
 \ifx #1\expandafter \@firstoftwo
 \else \expandafter \@secondoftwo
 \fi
}%
\providecommand \natexlab [1]{#1}%
\providecommand \enquote  [1]{``#1''}%
\providecommand \bibnamefont  [1]{#1}%
\providecommand \bibfnamefont [1]{#1}%
\providecommand \citenamefont [1]{#1}%
\providecommand \href@noop [0]{\@secondoftwo}%
\providecommand \href [0]{\begingroup \@sanitize@url \@href}%
\providecommand \@href[1]{\@@startlink{#1}\@@href}%
\providecommand \@@href[1]{\endgroup#1\@@endlink}%
\providecommand \@sanitize@url [0]{\catcode `\\12\catcode `\$12\catcode
  `\&12\catcode `\#12\catcode `\^12\catcode `\_12\catcode `\%12\relax}%
\providecommand \@@startlink[1]{}%
\providecommand \@@endlink[0]{}%
\providecommand \url  [0]{\begingroup\@sanitize@url \@url }%
\providecommand \@url [1]{\endgroup\@href {#1}{\urlprefix }}%
\providecommand \urlprefix  [0]{URL }%
\providecommand \Eprint [0]{\href }%
\providecommand \doibase [0]{https://doi.org/}%
\providecommand \selectlanguage [0]{\@gobble}%
\providecommand \bibinfo  [0]{\@secondoftwo}%
\providecommand \bibfield  [0]{\@secondoftwo}%
\providecommand \translation [1]{[#1]}%
\providecommand \BibitemOpen [0]{}%
\providecommand \bibitemStop [0]{}%
\providecommand \bibitemNoStop [0]{.\EOS\space}%
\providecommand \EOS [0]{\spacefactor3000\relax}%
\providecommand \BibitemShut  [1]{\csname bibitem#1\endcsname}%
\let\auto@bib@innerbib\@empty
\bibitem [{\citenamefont {Preskill}(2018)}]{computer1}%
  \BibitemOpen
  \bibfield  {author} {\bibinfo {author} {\bibfnamefont {J.}~\bibnamefont
  {Preskill}},\ }\bibfield  {title} {\bibinfo {title} {Quantum computing in the
  nisq era and beyond},\ }\href@noop {} {\bibfield  {journal} {\bibinfo
  {journal} {Quantum}\ }\textbf {\bibinfo {volume} {2}},\ \bibinfo {pages} {79}
  (\bibinfo {year} {2018})}\BibitemShut {NoStop}%
\bibitem [{\citenamefont {Ladd}\ \emph {et~al.}(2010)\citenamefont {Ladd},
  \citenamefont {Jelezko}, \citenamefont {Laflamme}, \citenamefont {Nakamura},
  \citenamefont {Monroe},\ and\ \citenamefont {O’Brien}}]{computer2}%
  \BibitemOpen
  \bibfield  {author} {\bibinfo {author} {\bibfnamefont {T.~D.}\ \bibnamefont
  {Ladd}}, \bibinfo {author} {\bibfnamefont {F.}~\bibnamefont {Jelezko}},
  \bibinfo {author} {\bibfnamefont {R.}~\bibnamefont {Laflamme}}, \bibinfo
  {author} {\bibfnamefont {Y.}~\bibnamefont {Nakamura}}, \bibinfo {author}
  {\bibfnamefont {C.}~\bibnamefont {Monroe}},\ and\ \bibinfo {author}
  {\bibfnamefont {J.~L.}\ \bibnamefont {O’Brien}},\ }\bibfield  {title}
  {\bibinfo {title} {Quantum computers},\ }\href@noop {} {\bibfield  {journal}
  {\bibinfo  {journal} {nature}\ }\textbf {\bibinfo {volume} {464}},\ \bibinfo
  {pages} {45} (\bibinfo {year} {2010})}\BibitemShut {NoStop}%
\bibitem [{\citenamefont {Nielsen}\ and\ \citenamefont
  {Chuang}(2010)}]{quantum1}%
  \BibitemOpen
  \bibfield  {author} {\bibinfo {author} {\bibfnamefont {M.~A.}\ \bibnamefont
  {Nielsen}}\ and\ \bibinfo {author} {\bibfnamefont {I.~L.}\ \bibnamefont
  {Chuang}},\ }\href@noop {} {\emph {\bibinfo {title} {Quantum computation and
  quantum information}}}\ (\bibinfo  {publisher} {Cambridge university press},\
  \bibinfo {year} {2010})\BibitemShut {NoStop}%
\bibitem [{\citenamefont {Feynman}(2018)}]{hai2}%
  \BibitemOpen
  \bibfield  {author} {\bibinfo {author} {\bibfnamefont {R.~P.}\ \bibnamefont
  {Feynman}},\ }\bibfield  {title} {\bibinfo {title} {Simulating physics with
  computers},\ }in\ \href@noop {} {\emph {\bibinfo {booktitle} {Feynman and
  computation}}}\ (\bibinfo  {publisher} {cRc Press},\ \bibinfo {year} {2018})\
  pp.\ \bibinfo {pages} {133--153}\BibitemShut {NoStop}%
\bibitem [{\citenamefont {Lloyd}(1996)}]{simulation3}%
  \BibitemOpen
  \bibfield  {author} {\bibinfo {author} {\bibfnamefont {S.}~\bibnamefont
  {Lloyd}},\ }\bibfield  {title} {\bibinfo {title} {Universal quantum
  simulators},\ }\href@noop {} {\bibfield  {journal} {\bibinfo  {journal}
  {Science}\ }\textbf {\bibinfo {volume} {273}},\ \bibinfo {pages} {1073}
  (\bibinfo {year} {1996})}\BibitemShut {NoStop}%
\bibitem [{\citenamefont {Vidal}(2003)}]{simulation4}%
  \BibitemOpen
  \bibfield  {author} {\bibinfo {author} {\bibfnamefont {G.}~\bibnamefont
  {Vidal}},\ }\bibfield  {title} {\bibinfo {title} {Efficient classical
  simulation of slightly entangled quantum computations},\ }\href@noop {}
  {\bibfield  {journal} {\bibinfo  {journal} {Physical review letters}\
  }\textbf {\bibinfo {volume} {91}},\ \bibinfo {pages} {147902} (\bibinfo
  {year} {2003})}\BibitemShut {NoStop}%
\bibitem [{\citenamefont {Georgescu}\ \emph {et~al.}(2014)\citenamefont
  {Georgescu}, \citenamefont {Ashhab},\ and\ \citenamefont
  {Nori}}]{simulation1}%
  \BibitemOpen
  \bibfield  {author} {\bibinfo {author} {\bibfnamefont {I.~M.}\ \bibnamefont
  {Georgescu}}, \bibinfo {author} {\bibfnamefont {S.}~\bibnamefont {Ashhab}},\
  and\ \bibinfo {author} {\bibfnamefont {F.}~\bibnamefont {Nori}},\ }\bibfield
  {title} {\bibinfo {title} {Quantum simulation},\ }\href@noop {} {\bibfield
  {journal} {\bibinfo  {journal} {Reviews of Modern Physics}\ }\textbf
  {\bibinfo {volume} {86}},\ \bibinfo {pages} {153} (\bibinfo {year}
  {2014})}\BibitemShut {NoStop}%
\bibitem [{\citenamefont {Kalyan}(2023)}]{gpt3}%
  \BibitemOpen
  \bibfield  {author} {\bibinfo {author} {\bibfnamefont {K.~S.}\ \bibnamefont
  {Kalyan}},\ }\bibfield  {title} {\bibinfo {title} {A survey of gpt-3 family
  large language models including chatgpt and gpt-4},\ }\href@noop {}
  {\bibfield  {journal} {\bibinfo  {journal} {Natural Language Processing
  Journal}\ ,\ \bibinfo {pages} {100048}} (\bibinfo {year} {2023})}\BibitemShut
  {NoStop}%
\bibitem [{\citenamefont {Brown}(2020)}]{llm8}%
  \BibitemOpen
  \bibfield  {author} {\bibinfo {author} {\bibfnamefont {T.~B.}\ \bibnamefont
  {Brown}},\ }\bibfield  {title} {\bibinfo {title} {Language models are
  few-shot learners},\ }\href@noop {} {\bibfield  {journal} {\bibinfo
  {journal} {arXiv preprint arXiv:2005.14165}\ } (\bibinfo {year}
  {2020})}\BibitemShut {NoStop}%
\bibitem [{\citenamefont {Sun}\ \emph {et~al.}(2019)\citenamefont {Sun},
  \citenamefont {Gaut}, \citenamefont {Tang}, \citenamefont {Huang},
  \citenamefont {ElSherief}, \citenamefont {Zhao}, \citenamefont {Mirza},
  \citenamefont {Belding}, \citenamefont {Chang},\ and\ \citenamefont
  {Wang}}]{language}%
  \BibitemOpen
  \bibfield  {author} {\bibinfo {author} {\bibfnamefont {T.}~\bibnamefont
  {Sun}}, \bibinfo {author} {\bibfnamefont {A.}~\bibnamefont {Gaut}}, \bibinfo
  {author} {\bibfnamefont {S.}~\bibnamefont {Tang}}, \bibinfo {author}
  {\bibfnamefont {Y.}~\bibnamefont {Huang}}, \bibinfo {author} {\bibfnamefont
  {M.}~\bibnamefont {ElSherief}}, \bibinfo {author} {\bibfnamefont
  {J.}~\bibnamefont {Zhao}}, \bibinfo {author} {\bibfnamefont {D.}~\bibnamefont
  {Mirza}}, \bibinfo {author} {\bibfnamefont {E.}~\bibnamefont {Belding}},
  \bibinfo {author} {\bibfnamefont {K.-W.}\ \bibnamefont {Chang}},\ and\
  \bibinfo {author} {\bibfnamefont {W.~Y.}\ \bibnamefont {Wang}},\ }\bibfield
  {title} {\bibinfo {title} {Mitigating gender bias in natural language
  processing: Literature review},\ }\href@noop {} {\bibfield  {journal}
  {\bibinfo  {journal} {arXiv preprint arXiv:1906.08976}\ } (\bibinfo {year}
  {2019})}\BibitemShut {NoStop}%
\bibitem [{\citenamefont {Cui}\ \emph {et~al.}(2020)\citenamefont {Cui},
  \citenamefont {Che}, \citenamefont {Liu}, \citenamefont {Qin}, \citenamefont
  {Wang},\ and\ \citenamefont {Hu}}]{language2}%
  \BibitemOpen
  \bibfield  {author} {\bibinfo {author} {\bibfnamefont {Y.}~\bibnamefont
  {Cui}}, \bibinfo {author} {\bibfnamefont {W.}~\bibnamefont {Che}}, \bibinfo
  {author} {\bibfnamefont {T.}~\bibnamefont {Liu}}, \bibinfo {author}
  {\bibfnamefont {B.}~\bibnamefont {Qin}}, \bibinfo {author} {\bibfnamefont
  {S.}~\bibnamefont {Wang}},\ and\ \bibinfo {author} {\bibfnamefont
  {G.}~\bibnamefont {Hu}},\ }\bibfield  {title} {\bibinfo {title} {Revisiting
  pre-trained models for chinese natural language processing},\ }\href@noop {}
  {\bibfield  {journal} {\bibinfo  {journal} {arXiv preprint arXiv:2004.13922}\
  } (\bibinfo {year} {2020})}\BibitemShut {NoStop}%
\bibitem [{\citenamefont {Goldberg}(2022)}]{nlp3}%
  \BibitemOpen
  \bibfield  {author} {\bibinfo {author} {\bibfnamefont {Y.}~\bibnamefont
  {Goldberg}},\ }\href@noop {} {\emph {\bibinfo {title} {Neural network methods
  for natural language processing}}}\ (\bibinfo  {publisher} {Springer
  Nature},\ \bibinfo {year} {2022})\BibitemShut {NoStop}%
\bibitem [{\citenamefont {Esteva}\ \emph {et~al.}(2019)\citenamefont {Esteva},
  \citenamefont {Robicquet}, \citenamefont {Ramsundar}, \citenamefont
  {Kuleshov}, \citenamefont {DePristo}, \citenamefont {Chou}, \citenamefont
  {Cui}, \citenamefont {Corrado}, \citenamefont {Thrun},\ and\ \citenamefont
  {Dean}}]{nlp6}%
  \BibitemOpen
  \bibfield  {author} {\bibinfo {author} {\bibfnamefont {A.}~\bibnamefont
  {Esteva}}, \bibinfo {author} {\bibfnamefont {A.}~\bibnamefont {Robicquet}},
  \bibinfo {author} {\bibfnamefont {B.}~\bibnamefont {Ramsundar}}, \bibinfo
  {author} {\bibfnamefont {V.}~\bibnamefont {Kuleshov}}, \bibinfo {author}
  {\bibfnamefont {M.}~\bibnamefont {DePristo}}, \bibinfo {author}
  {\bibfnamefont {K.}~\bibnamefont {Chou}}, \bibinfo {author} {\bibfnamefont
  {C.}~\bibnamefont {Cui}}, \bibinfo {author} {\bibfnamefont {G.}~\bibnamefont
  {Corrado}}, \bibinfo {author} {\bibfnamefont {S.}~\bibnamefont {Thrun}},\
  and\ \bibinfo {author} {\bibfnamefont {J.}~\bibnamefont {Dean}},\ }\bibfield
  {title} {\bibinfo {title} {A guide to deep learning in healthcare},\
  }\href@noop {} {\bibfield  {journal} {\bibinfo  {journal} {Nature medicine}\
  }\textbf {\bibinfo {volume} {25}},\ \bibinfo {pages} {24} (\bibinfo {year}
  {2019})}\BibitemShut {NoStop}%
\bibitem [{\citenamefont {Schuld}\ and\ \citenamefont
  {Petruccione}(2021)}]{llm2}%
  \BibitemOpen
  \bibfield  {author} {\bibinfo {author} {\bibfnamefont {M.}~\bibnamefont
  {Schuld}}\ and\ \bibinfo {author} {\bibfnamefont {F.}~\bibnamefont
  {Petruccione}},\ }\href@noop {} {\emph {\bibinfo {title} {Machine learning
  with quantum computers}}},\ Vol.\ \bibinfo {volume} {676}\ (\bibinfo
  {publisher} {Springer},\ \bibinfo {year} {2021})\BibitemShut {NoStop}%
\bibitem [{\citenamefont {Kawai}\ and\ \citenamefont {Nakagawa}(2020)}]{llm3}%
  \BibitemOpen
  \bibfield  {author} {\bibinfo {author} {\bibfnamefont {H.}~\bibnamefont
  {Kawai}}\ and\ \bibinfo {author} {\bibfnamefont {Y.~O.}\ \bibnamefont
  {Nakagawa}},\ }\bibfield  {title} {\bibinfo {title} {Predicting excited
  states from ground state wavefunction by supervised quantum machine
  learning},\ }\href@noop {} {\bibfield  {journal} {\bibinfo  {journal}
  {Machine Learning: Science and Technology}\ }\textbf {\bibinfo {volume}
  {1}},\ \bibinfo {pages} {045027} (\bibinfo {year} {2020})}\BibitemShut
  {NoStop}%
\bibitem [{\citenamefont {Kharsa}\ \emph {et~al.}(2023)\citenamefont {Kharsa},
  \citenamefont {Bouridane},\ and\ \citenamefont {Amira}}]{fenl}%
  \BibitemOpen
  \bibfield  {author} {\bibinfo {author} {\bibfnamefont {R.}~\bibnamefont
  {Kharsa}}, \bibinfo {author} {\bibfnamefont {A.}~\bibnamefont {Bouridane}},\
  and\ \bibinfo {author} {\bibfnamefont {A.}~\bibnamefont {Amira}},\ }\bibfield
   {title} {\bibinfo {title} {Advances in quantum machine learning and deep
  learning for image classification: a survey},\ }\href@noop {} {\bibfield
  {journal} {\bibinfo  {journal} {Neurocomputing}\ }\textbf {\bibinfo {volume}
  {560}},\ \bibinfo {pages} {126843} (\bibinfo {year} {2023})}\BibitemShut
  {NoStop}%
\bibitem [{\citenamefont {Abohashima}\ \emph {et~al.}(2020)\citenamefont
  {Abohashima}, \citenamefont {Elhosen}, \citenamefont {Houssein},\ and\
  \citenamefont {Mohamed}}]{fenl2}%
  \BibitemOpen
  \bibfield  {author} {\bibinfo {author} {\bibfnamefont {Z.}~\bibnamefont
  {Abohashima}}, \bibinfo {author} {\bibfnamefont {M.}~\bibnamefont {Elhosen}},
  \bibinfo {author} {\bibfnamefont {E.~H.}\ \bibnamefont {Houssein}},\ and\
  \bibinfo {author} {\bibfnamefont {W.~M.}\ \bibnamefont {Mohamed}},\
  }\bibfield  {title} {\bibinfo {title} {Classification with quantum machine
  learning: A survey},\ }\href@noop {} {\bibfield  {journal} {\bibinfo
  {journal} {arXiv preprint arXiv:2006.12270}\ } (\bibinfo {year}
  {2020})}\BibitemShut {NoStop}%
\bibitem [{\citenamefont {Radford}\ \emph {et~al.}(2019)\citenamefont
  {Radford}, \citenamefont {Wu}, \citenamefont {Child}, \citenamefont {Luan},
  \citenamefont {Amodei}, \citenamefont {Sutskever} \emph {et~al.}}]{llm7}%
  \BibitemOpen
  \bibfield  {author} {\bibinfo {author} {\bibfnamefont {A.}~\bibnamefont
  {Radford}}, \bibinfo {author} {\bibfnamefont {J.}~\bibnamefont {Wu}},
  \bibinfo {author} {\bibfnamefont {R.}~\bibnamefont {Child}}, \bibinfo
  {author} {\bibfnamefont {D.}~\bibnamefont {Luan}}, \bibinfo {author}
  {\bibfnamefont {D.}~\bibnamefont {Amodei}}, \bibinfo {author} {\bibfnamefont
  {I.}~\bibnamefont {Sutskever}}, \emph {et~al.},\ }\bibfield  {title}
  {\bibinfo {title} {Language models are unsupervised multitask learners},\
  }\href@noop {} {\bibfield  {journal} {\bibinfo  {journal} {OpenAI blog}\
  }\textbf {\bibinfo {volume} {1}},\ \bibinfo {pages} {9} (\bibinfo {year}
  {2019})}\BibitemShut {NoStop}%
\bibitem [{\citenamefont {Devlin}(2018)}]{zhuc1}%
  \BibitemOpen
  \bibfield  {author} {\bibinfo {author} {\bibfnamefont {J.}~\bibnamefont
  {Devlin}},\ }\bibfield  {title} {\bibinfo {title} {Bert: Pre-training of deep
  bidirectional transformers for language understanding},\ }\href@noop {}
  {\bibfield  {journal} {\bibinfo  {journal} {arXiv preprint arXiv:1810.04805}\
  } (\bibinfo {year} {2018})}\BibitemShut {NoStop}%
\bibitem [{\citenamefont {Vaswani}(2017)}]{attention1}%
  \BibitemOpen
  \bibfield  {author} {\bibinfo {author} {\bibfnamefont {A.}~\bibnamefont
  {Vaswani}},\ }\bibfield  {title} {\bibinfo {title} {Attention is all you
  need},\ }\href@noop {} {\bibfield  {journal} {\bibinfo  {journal} {Advances
  in Neural Information Processing Systems}\ } (\bibinfo {year}
  {2017})}\BibitemShut {NoStop}%
\bibitem [{\citenamefont {Barenco}\ \emph {et~al.}(1995)\citenamefont
  {Barenco}, \citenamefont {Bennett}, \citenamefont {Cleve}, \citenamefont
  {DiVincenzo}, \citenamefont {Margolus}, \citenamefont {Shor}, \citenamefont
  {Sleator}, \citenamefont {Smolin},\ and\ \citenamefont
  {Weinfurter}}]{quantum2}%
  \BibitemOpen
  \bibfield  {author} {\bibinfo {author} {\bibfnamefont {A.}~\bibnamefont
  {Barenco}}, \bibinfo {author} {\bibfnamefont {C.~H.}\ \bibnamefont
  {Bennett}}, \bibinfo {author} {\bibfnamefont {R.}~\bibnamefont {Cleve}},
  \bibinfo {author} {\bibfnamefont {D.~P.}\ \bibnamefont {DiVincenzo}},
  \bibinfo {author} {\bibfnamefont {N.}~\bibnamefont {Margolus}}, \bibinfo
  {author} {\bibfnamefont {P.}~\bibnamefont {Shor}}, \bibinfo {author}
  {\bibfnamefont {T.}~\bibnamefont {Sleator}}, \bibinfo {author} {\bibfnamefont
  {J.~A.}\ \bibnamefont {Smolin}},\ and\ \bibinfo {author} {\bibfnamefont
  {H.}~\bibnamefont {Weinfurter}},\ }\bibfield  {title} {\bibinfo {title}
  {Elementary gates for quantum computation},\ }\href@noop {} {\bibfield
  {journal} {\bibinfo  {journal} {Physical review A}\ }\textbf {\bibinfo
  {volume} {52}},\ \bibinfo {pages} {3457} (\bibinfo {year}
  {1995})}\BibitemShut {NoStop}%
\bibitem [{\citenamefont {Jozsa}(1994)}]{fid1}%
  \BibitemOpen
  \bibfield  {author} {\bibinfo {author} {\bibfnamefont {R.}~\bibnamefont
  {Jozsa}},\ }\bibfield  {title} {\bibinfo {title} {Fidelity for mixed quantum
  states},\ }\href@noop {} {\bibfield  {journal} {\bibinfo  {journal} {Journal
  of modern optics}\ }\textbf {\bibinfo {volume} {41}},\ \bibinfo {pages}
  {2315} (\bibinfo {year} {1994})}\BibitemShut {NoStop}%
\bibitem [{\citenamefont {Bengtsson}\ and\ \citenamefont
  {{\.Z}yczkowski}(2017{\natexlab{a}})}]{fid2}%
  \BibitemOpen
  \bibfield  {author} {\bibinfo {author} {\bibfnamefont {I.}~\bibnamefont
  {Bengtsson}}\ and\ \bibinfo {author} {\bibfnamefont {K.}~\bibnamefont
  {{\.Z}yczkowski}},\ }\href@noop {} {\emph {\bibinfo {title} {Geometry of
  quantum states: an introduction to quantum entanglement}}}\ (\bibinfo
  {publisher} {Cambridge university press},\ \bibinfo {year}
  {2017})\BibitemShut {NoStop}%
\bibitem [{\citenamefont {Uhlmann}(1976)}]{fid3}%
  \BibitemOpen
  \bibfield  {author} {\bibinfo {author} {\bibfnamefont {A.}~\bibnamefont
  {Uhlmann}},\ }\bibfield  {title} {\bibinfo {title} {The “transition
  probability” in the state space of a*-algebra},\ }\href@noop {} {\bibfield
  {journal} {\bibinfo  {journal} {Reports on Mathematical Physics}\ }\textbf
  {\bibinfo {volume} {9}},\ \bibinfo {pages} {273} (\bibinfo {year}
  {1976})}\BibitemShut {NoStop}%
\bibitem [{\citenamefont {Mermin}(2006)}]{quantum4}%
  \BibitemOpen
  \bibfield  {author} {\bibinfo {author} {\bibfnamefont {N.~D.}\ \bibnamefont
  {Mermin}},\ }\bibfield  {title} {\bibinfo {title} {Lecture notes on quantum
  computation},\ }\href@noop {} {\bibfield  {journal} {\bibinfo  {journal}
  {Cornell University}\ } (\bibinfo {year} {2006})}\BibitemShut {NoStop}%
\bibitem [{\citenamefont {Breuer}\ and\ \citenamefont
  {Petruccione}(2002)}]{quantum3}%
  \BibitemOpen
  \bibfield  {author} {\bibinfo {author} {\bibfnamefont {H.-P.}\ \bibnamefont
  {Breuer}}\ and\ \bibinfo {author} {\bibfnamefont {F.}~\bibnamefont
  {Petruccione}},\ }\href@noop {} {\emph {\bibinfo {title} {The theory of open
  quantum systems}}}\ (\bibinfo  {publisher} {Oxford University Press, USA},\
  \bibinfo {year} {2002})\BibitemShut {NoStop}%
\bibitem [{\citenamefont {Peruzzo}\ \emph {et~al.}(2013)\citenamefont {Peruzzo}
  \emph {et~al.}}]{vqe1}%
  \BibitemOpen
  \bibfield  {author} {\bibinfo {author} {\bibfnamefont {A.}~\bibnamefont
  {Peruzzo}} \emph {et~al.},\ }\bibfield  {title} {\bibinfo {title} {A
  variational eigenvalue solver on a quantum processor. eprint},\ }\href@noop
  {} {\bibfield  {journal} {\bibinfo  {journal} {arXiv preprint
  arXiv:1304.3061}\ } (\bibinfo {year} {2013})}\BibitemShut {NoStop}%
\bibitem [{\citenamefont {Cerezo}\ \emph {et~al.}(2021)\citenamefont {Cerezo},
  \citenamefont {Arrasmith}, \citenamefont {Babbush}, \citenamefont {Benjamin},
  \citenamefont {Endo}, \citenamefont {Fujii}, \citenamefont {McClean},
  \citenamefont {Mitarai}, \citenamefont {Yuan}, \citenamefont {Cincio} \emph
  {et~al.}}]{vqe2}%
  \BibitemOpen
  \bibfield  {author} {\bibinfo {author} {\bibfnamefont {M.}~\bibnamefont
  {Cerezo}}, \bibinfo {author} {\bibfnamefont {A.}~\bibnamefont {Arrasmith}},
  \bibinfo {author} {\bibfnamefont {R.}~\bibnamefont {Babbush}}, \bibinfo
  {author} {\bibfnamefont {S.~C.}\ \bibnamefont {Benjamin}}, \bibinfo {author}
  {\bibfnamefont {S.}~\bibnamefont {Endo}}, \bibinfo {author} {\bibfnamefont
  {K.}~\bibnamefont {Fujii}}, \bibinfo {author} {\bibfnamefont {J.~R.}\
  \bibnamefont {McClean}}, \bibinfo {author} {\bibfnamefont {K.}~\bibnamefont
  {Mitarai}}, \bibinfo {author} {\bibfnamefont {X.}~\bibnamefont {Yuan}},
  \bibinfo {author} {\bibfnamefont {L.}~\bibnamefont {Cincio}}, \emph
  {et~al.},\ }\bibfield  {title} {\bibinfo {title} {Variational quantum
  algorithms},\ }\href@noop {} {\bibfield  {journal} {\bibinfo  {journal}
  {Nature Reviews Physics}\ }\textbf {\bibinfo {volume} {3}},\ \bibinfo {pages}
  {625} (\bibinfo {year} {2021})}\BibitemShut {NoStop}%
\bibitem [{\citenamefont {Auerbach}(2012)}]{hai1}%
  \BibitemOpen
  \bibfield  {author} {\bibinfo {author} {\bibfnamefont {A.}~\bibnamefont
  {Auerbach}},\ }\href@noop {} {\emph {\bibinfo {title} {Interacting electrons
  and quantum magnetism}}}\ (\bibinfo  {publisher} {Springer Science \&
  Business Media},\ \bibinfo {year} {2012})\BibitemShut {NoStop}%
\bibitem [{\citenamefont {Knill}(2005)}]{hai3}%
  \BibitemOpen
  \bibfield  {author} {\bibinfo {author} {\bibfnamefont {E.}~\bibnamefont
  {Knill}},\ }\bibfield  {title} {\bibinfo {title} {Quantum computing with
  realistically noisy devices},\ }\href@noop {} {\bibfield  {journal} {\bibinfo
   {journal} {Nature}\ }\textbf {\bibinfo {volume} {434}},\ \bibinfo {pages}
  {39} (\bibinfo {year} {2005})}\BibitemShut {NoStop}%
\bibitem [{\citenamefont {Aleksandrowicz}\ \emph {et~al.}(2019)\citenamefont
  {Aleksandrowicz}, \citenamefont {Alexander}, \citenamefont {Barkoutsos},
  \citenamefont {Bello}, \citenamefont {Ben-Haim}, \citenamefont {Bucher},
  \citenamefont {Cabrera-Hern{\'a}ndez}, \citenamefont {Carballo-Franquis},
  \citenamefont {Chen}, \citenamefont {Chen} \emph {et~al.}}]{qiskit1}%
  \BibitemOpen
  \bibfield  {author} {\bibinfo {author} {\bibfnamefont {G.}~\bibnamefont
  {Aleksandrowicz}}, \bibinfo {author} {\bibfnamefont {T.}~\bibnamefont
  {Alexander}}, \bibinfo {author} {\bibfnamefont {P.}~\bibnamefont
  {Barkoutsos}}, \bibinfo {author} {\bibfnamefont {L.}~\bibnamefont {Bello}},
  \bibinfo {author} {\bibfnamefont {Y.}~\bibnamefont {Ben-Haim}}, \bibinfo
  {author} {\bibfnamefont {D.}~\bibnamefont {Bucher}}, \bibinfo {author}
  {\bibfnamefont {F.~J.}\ \bibnamefont {Cabrera-Hern{\'a}ndez}}, \bibinfo
  {author} {\bibfnamefont {J.}~\bibnamefont {Carballo-Franquis}}, \bibinfo
  {author} {\bibfnamefont {A.}~\bibnamefont {Chen}}, \bibinfo {author}
  {\bibfnamefont {C.-F.}\ \bibnamefont {Chen}}, \emph {et~al.},\ }\bibfield
  {title} {\bibinfo {title} {Qiskit: An open-source framework for quantum
  computing},\ }\href@noop {} {\bibfield  {journal} {\bibinfo  {journal}
  {Accessed on: Mar}\ }\textbf {\bibinfo {volume} {16}},\ \bibinfo {pages} {61}
  (\bibinfo {year} {2019})}\BibitemShut {NoStop}%
\bibitem [{\citenamefont {Jones}\ and\ \citenamefont {Gacon}(2020)}]{qiskit2}%
  \BibitemOpen
  \bibfield  {author} {\bibinfo {author} {\bibfnamefont {T.}~\bibnamefont
  {Jones}}\ and\ \bibinfo {author} {\bibfnamefont {J.}~\bibnamefont {Gacon}},\
  }\bibfield  {title} {\bibinfo {title} {Efficient calculation of gradients in
  classical simulations of variational quantum algorithms},\ }\href@noop {}
  {\bibfield  {journal} {\bibinfo  {journal} {arXiv preprint arXiv:2009.02823}\
  } (\bibinfo {year} {2020})}\BibitemShut {NoStop}%
\bibitem [{\citenamefont {Suh}\ and\ \citenamefont {Li}(2024)}]{qiskit3}%
  \BibitemOpen
  \bibfield  {author} {\bibinfo {author} {\bibfnamefont {I.-S.}\ \bibnamefont
  {Suh}}\ and\ \bibinfo {author} {\bibfnamefont {A.}~\bibnamefont {Li}},\
  }\bibfield  {title} {\bibinfo {title} {Simulating quantum systems with
  nwq-sim on hpc},\ }\href@noop {} {\bibfield  {journal} {\bibinfo  {journal}
  {arXiv preprint arXiv:2401.06861}\ } (\bibinfo {year} {2024})}\BibitemShut
  {NoStop}%
\bibitem [{\citenamefont {{\"O}berg}\ and\ \citenamefont
  {Shahriari}(2023)}]{qiskit4}%
  \BibitemOpen
  \bibfield  {author} {\bibinfo {author} {\bibfnamefont {W.}~\bibnamefont
  {{\"O}berg}}\ and\ \bibinfo {author} {\bibfnamefont {S.}~\bibnamefont
  {Shahriari}},\ }\href@noop {} {\bibinfo {title} {Simulating the impact of
  noise on quantum walk algorithm}} (\bibinfo {year} {2023})\BibitemShut
  {NoStop}%
\bibitem [{\citenamefont {Kandala}\ \emph {et~al.}(2017)\citenamefont
  {Kandala}, \citenamefont {Mezzacapo}, \citenamefont {Temme}, \citenamefont
  {Takita}, \citenamefont {Brink}, \citenamefont {Chow},\ and\ \citenamefont
  {Gambetta}}]{vqe5}%
  \BibitemOpen
  \bibfield  {author} {\bibinfo {author} {\bibfnamefont {A.}~\bibnamefont
  {Kandala}}, \bibinfo {author} {\bibfnamefont {A.}~\bibnamefont {Mezzacapo}},
  \bibinfo {author} {\bibfnamefont {K.}~\bibnamefont {Temme}}, \bibinfo
  {author} {\bibfnamefont {M.}~\bibnamefont {Takita}}, \bibinfo {author}
  {\bibfnamefont {M.}~\bibnamefont {Brink}}, \bibinfo {author} {\bibfnamefont
  {J.~M.}\ \bibnamefont {Chow}},\ and\ \bibinfo {author} {\bibfnamefont
  {J.~M.}\ \bibnamefont {Gambetta}},\ }\bibfield  {title} {\bibinfo {title}
  {Hardware-efficient variational quantum eigensolver for small molecules and
  quantum magnets},\ }\href@noop {} {\bibfield  {journal} {\bibinfo  {journal}
  {nature}\ }\textbf {\bibinfo {volume} {549}},\ \bibinfo {pages} {242}
  (\bibinfo {year} {2017})}\BibitemShut {NoStop}%
\bibitem [{\citenamefont {McArdle}\ \emph {et~al.}(2020)\citenamefont
  {McArdle}, \citenamefont {Endo}, \citenamefont {Aspuru-Guzik}, \citenamefont
  {Benjamin},\ and\ \citenamefont {Yuan}}]{vqe6}%
  \BibitemOpen
  \bibfield  {author} {\bibinfo {author} {\bibfnamefont {S.}~\bibnamefont
  {McArdle}}, \bibinfo {author} {\bibfnamefont {S.}~\bibnamefont {Endo}},
  \bibinfo {author} {\bibfnamefont {A.}~\bibnamefont {Aspuru-Guzik}}, \bibinfo
  {author} {\bibfnamefont {S.~C.}\ \bibnamefont {Benjamin}},\ and\ \bibinfo
  {author} {\bibfnamefont {X.}~\bibnamefont {Yuan}},\ }\bibfield  {title}
  {\bibinfo {title} {Quantum computational chemistry},\ }\href@noop {}
  {\bibfield  {journal} {\bibinfo  {journal} {Reviews of Modern Physics}\
  }\textbf {\bibinfo {volume} {92}},\ \bibinfo {pages} {015003} (\bibinfo
  {year} {2020})}\BibitemShut {NoStop}%
\bibitem [{\citenamefont {Grover}(1996)}]{grover1}%
  \BibitemOpen
  \bibfield  {author} {\bibinfo {author} {\bibfnamefont {L.~K.}\ \bibnamefont
  {Grover}},\ }\bibfield  {title} {\bibinfo {title} {A fast quantum mechanical
  algorithm for database search},\ }in\ \href@noop {} {\emph {\bibinfo
  {booktitle} {Proceedings of the twenty-eighth annual ACM symposium on Theory
  of computing}}}\ (\bibinfo {year} {1996})\ pp.\ \bibinfo {pages}
  {212--219}\BibitemShut {NoStop}%
\bibitem [{\citenamefont {Zalka}(1999)}]{grover2}%
  \BibitemOpen
  \bibfield  {author} {\bibinfo {author} {\bibfnamefont {C.}~\bibnamefont
  {Zalka}},\ }\bibfield  {title} {\bibinfo {title} {Grover’s quantum
  searching algorithm is optimal},\ }\href@noop {} {\bibfield  {journal}
  {\bibinfo  {journal} {Physical Review A}\ }\textbf {\bibinfo {volume} {60}},\
  \bibinfo {pages} {2746} (\bibinfo {year} {1999})}\BibitemShut {NoStop}%
\bibitem [{\citenamefont {Long}(2001)}]{grover3}%
  \BibitemOpen
  \bibfield  {author} {\bibinfo {author} {\bibfnamefont {G.-L.}\ \bibnamefont
  {Long}},\ }\bibfield  {title} {\bibinfo {title} {Grover algorithm with zero
  theoretical failure rate},\ }\href@noop {} {\bibfield  {journal} {\bibinfo
  {journal} {Physical Review A}\ }\textbf {\bibinfo {volume} {64}},\ \bibinfo
  {pages} {022307} (\bibinfo {year} {2001})}\BibitemShut {NoStop}%
\bibitem [{\citenamefont {Peres}(1997)}]{theory1}%
  \BibitemOpen
  \bibfield  {author} {\bibinfo {author} {\bibfnamefont {A.}~\bibnamefont
  {Peres}},\ }\href@noop {} {\emph {\bibinfo {title} {Quantum theory: concepts
  and methods}}},\ Vol.~\bibinfo {volume} {72}\ (\bibinfo  {publisher}
  {Springer},\ \bibinfo {year} {1997})\BibitemShut {NoStop}%
\bibitem [{\citenamefont {Dirac}(1981)}]{theory2}%
  \BibitemOpen
  \bibfield  {author} {\bibinfo {author} {\bibfnamefont {P.~A.~M.}\
  \bibnamefont {Dirac}},\ }\href@noop {} {\emph {\bibinfo {title} {The
  principles of quantum mechanics}}},\ \bibinfo {number} {27}\ (\bibinfo
  {publisher} {Oxford university press},\ \bibinfo {year} {1981})\BibitemShut
  {NoStop}%
\bibitem [{\citenamefont {Horn}\ and\ \citenamefont {Johnson}(2012)}]{theory3}%
  \BibitemOpen
  \bibfield  {author} {\bibinfo {author} {\bibfnamefont {R.~A.}\ \bibnamefont
  {Horn}}\ and\ \bibinfo {author} {\bibfnamefont {C.~R.}\ \bibnamefont
  {Johnson}},\ }\href@noop {} {\emph {\bibinfo {title} {Matrix analysis}}}\
  (\bibinfo  {publisher} {Cambridge university press},\ \bibinfo {year}
  {2012})\BibitemShut {NoStop}%
\bibitem [{\citenamefont {Ithier}\ \emph {et~al.}(2005)\citenamefont {Ithier},
  \citenamefont {Collin}, \citenamefont {Joyez}, \citenamefont {Meeson},
  \citenamefont {Vion}, \citenamefont {Esteve}, \citenamefont {Chiarello},
  \citenamefont {Shnirman}, \citenamefont {Makhlin}, \citenamefont {Schriefl}
  \emph {et~al.}}]{noise}%
  \BibitemOpen
  \bibfield  {author} {\bibinfo {author} {\bibfnamefont {G.}~\bibnamefont
  {Ithier}}, \bibinfo {author} {\bibfnamefont {E.}~\bibnamefont {Collin}},
  \bibinfo {author} {\bibfnamefont {P.}~\bibnamefont {Joyez}}, \bibinfo
  {author} {\bibfnamefont {P.}~\bibnamefont {Meeson}}, \bibinfo {author}
  {\bibfnamefont {D.}~\bibnamefont {Vion}}, \bibinfo {author} {\bibfnamefont
  {D.}~\bibnamefont {Esteve}}, \bibinfo {author} {\bibfnamefont
  {F.}~\bibnamefont {Chiarello}}, \bibinfo {author} {\bibfnamefont
  {A.}~\bibnamefont {Shnirman}}, \bibinfo {author} {\bibfnamefont
  {Y.}~\bibnamefont {Makhlin}}, \bibinfo {author} {\bibfnamefont
  {J.}~\bibnamefont {Schriefl}}, \emph {et~al.},\ }\bibfield  {title} {\bibinfo
  {title} {Decoherence in a superconducting quantum bit circuit},\ }\href@noop
  {} {\bibfield  {journal} {\bibinfo  {journal} {Physical Review B—Condensed
  Matter and Materials Physics}\ }\textbf {\bibinfo {volume} {72}},\ \bibinfo
  {pages} {134519} (\bibinfo {year} {2005})}\BibitemShut {NoStop}%
\bibitem [{\citenamefont {Lee}(2002)}]{lee}%
  \BibitemOpen
  \bibfield  {author} {\bibinfo {author} {\bibfnamefont {J.-S.}\ \bibnamefont
  {Lee}},\ }\bibfield  {title} {\bibinfo {title} {The quantum state tomography
  on an nmr system},\ }\href@noop {} {\bibfield  {journal} {\bibinfo  {journal}
  {Physics Letters A}\ }\textbf {\bibinfo {volume} {305}},\ \bibinfo {pages}
  {349} (\bibinfo {year} {2002})}\BibitemShut {NoStop}%
\bibitem [{\citenamefont {Long}\ \emph {et~al.}(2001)\citenamefont {Long},
  \citenamefont {Yan},\ and\ \citenamefont {Sun}}]{long1}%
  \BibitemOpen
  \bibfield  {author} {\bibinfo {author} {\bibfnamefont {G.}~\bibnamefont
  {Long}}, \bibinfo {author} {\bibfnamefont {H.}~\bibnamefont {Yan}},\ and\
  \bibinfo {author} {\bibfnamefont {Y.}~\bibnamefont {Sun}},\ }\bibfield
  {title} {\bibinfo {title} {Analysis of density matrix reconstruction in nmr
  quantum computing},\ }\href@noop {} {\bibfield  {journal} {\bibinfo
  {journal} {Journal of Optics B: quantum and semiclassical optics}\ }\textbf
  {\bibinfo {volume} {3}},\ \bibinfo {pages} {376} (\bibinfo {year}
  {2001})}\BibitemShut {NoStop}%
\bibitem [{\citenamefont {Bengtsson}\ and\ \citenamefont
  {{\.Z}yczkowski}(2017{\natexlab{b}})}]{convex5}%
  \BibitemOpen
  \bibfield  {author} {\bibinfo {author} {\bibfnamefont {I.}~\bibnamefont
  {Bengtsson}}\ and\ \bibinfo {author} {\bibfnamefont {K.}~\bibnamefont
  {{\.Z}yczkowski}},\ }\href@noop {} {\emph {\bibinfo {title} {Geometry of
  quantum states: an introduction to quantum entanglement}}}\ (\bibinfo
  {publisher} {Cambridge university press},\ \bibinfo {year}
  {2017})\BibitemShut {NoStop}%
\bibitem [{\citenamefont {Vogel}(1990)}]{vogel1990constrained}%
  \BibitemOpen
  \bibfield  {author} {\bibinfo {author} {\bibfnamefont {C.~R.}\ \bibnamefont
  {Vogel}},\ }\bibfield  {title} {\bibinfo {title} {A constrained least squares
  regularization method for nonlinear iii-posed problems},\ }\href@noop {}
  {\bibfield  {journal} {\bibinfo  {journal} {SIAM Journal on Control and
  Optimization}\ }\textbf {\bibinfo {volume} {28}},\ \bibinfo {pages} {34}
  (\bibinfo {year} {1990})}\BibitemShut {NoStop}%
\bibitem [{\citenamefont {Kirsch}\ \emph {et~al.}(2011)\citenamefont {Kirsch}
  \emph {et~al.}}]{kirsch2011introduction}%
  \BibitemOpen
  \bibfield  {author} {\bibinfo {author} {\bibfnamefont {A.}~\bibnamefont
  {Kirsch}} \emph {et~al.},\ }\href@noop {} {\emph {\bibinfo {title} {An
  introduction to the mathematical theory of inverse problems}}},\ Vol.\
  \bibinfo {volume} {120}\ (\bibinfo  {publisher} {Springer},\ \bibinfo {year}
  {2011})\BibitemShut {NoStop}%
\bibitem [{\citenamefont {Souopgui}\ \emph {et~al.}(2016)\citenamefont
  {Souopgui}, \citenamefont {Ngodock}, \citenamefont {Vidard},\ and\
  \citenamefont {Le~Dimet}}]{souopgui2016incremental}%
  \BibitemOpen
  \bibfield  {author} {\bibinfo {author} {\bibfnamefont {I.}~\bibnamefont
  {Souopgui}}, \bibinfo {author} {\bibfnamefont {H.~E.}\ \bibnamefont
  {Ngodock}}, \bibinfo {author} {\bibfnamefont {A.}~\bibnamefont {Vidard}},\
  and\ \bibinfo {author} {\bibfnamefont {F.-X.}\ \bibnamefont {Le~Dimet}},\
  }\bibfield  {title} {\bibinfo {title} {Incremental projection approach of
  regularization for inverse problems},\ }\href@noop {} {\bibfield  {journal}
  {\bibinfo  {journal} {Applied Mathematics \& Optimization}\ }\textbf
  {\bibinfo {volume} {74}},\ \bibinfo {pages} {303} (\bibinfo {year}
  {2016})}\BibitemShut {NoStop}%
\bibitem [{\citenamefont {Boyd}\ and\ \citenamefont
  {Vandenberghe}(2004)}]{convex4}%
  \BibitemOpen
  \bibfield  {author} {\bibinfo {author} {\bibfnamefont {S.}~\bibnamefont
  {Boyd}}\ and\ \bibinfo {author} {\bibfnamefont {L.}~\bibnamefont
  {Vandenberghe}},\ }\href@noop {} {\emph {\bibinfo {title} {Convex
  optimization}}}\ (\bibinfo  {publisher} {Cambridge university press},\
  \bibinfo {year} {2004})\BibitemShut {NoStop}%
\bibitem [{\citenamefont {Diamond}\ and\ \citenamefont {Boyd}(2016)}]{cvxpy1}%
  \BibitemOpen
  \bibfield  {author} {\bibinfo {author} {\bibfnamefont {S.}~\bibnamefont
  {Diamond}}\ and\ \bibinfo {author} {\bibfnamefont {S.}~\bibnamefont {Boyd}},\
  }\bibfield  {title} {\bibinfo {title} {Cvxpy: A python-embedded modeling
  language for convex optimization},\ }\href@noop {} {\bibfield  {journal}
  {\bibinfo  {journal} {Journal of Machine Learning Research}\ }\textbf
  {\bibinfo {volume} {17}},\ \bibinfo {pages} {1} (\bibinfo {year}
  {2016})}\BibitemShut {NoStop}%
\bibitem [{\citenamefont {Strandberg}(2022)}]{cvxpy2}%
  \BibitemOpen
  \bibfield  {author} {\bibinfo {author} {\bibfnamefont {I.}~\bibnamefont
  {Strandberg}},\ }\bibfield  {title} {\bibinfo {title} {Simple, reliable, and
  noise-resilient continuous-variable quantum state tomography with convex
  optimization},\ }\href@noop {} {\bibfield  {journal} {\bibinfo  {journal}
  {Physical Review Applied}\ }\textbf {\bibinfo {volume} {18}},\ \bibinfo
  {pages} {044041} (\bibinfo {year} {2022})}\BibitemShut {NoStop}%
\bibitem [{\citenamefont {Agrawal}\ \emph {et~al.}(2018)\citenamefont
  {Agrawal}, \citenamefont {Verschueren}, \citenamefont {Diamond},\ and\
  \citenamefont {Boyd}}]{cvxpy3}%
  \BibitemOpen
  \bibfield  {author} {\bibinfo {author} {\bibfnamefont {A.}~\bibnamefont
  {Agrawal}}, \bibinfo {author} {\bibfnamefont {R.}~\bibnamefont
  {Verschueren}}, \bibinfo {author} {\bibfnamefont {S.}~\bibnamefont
  {Diamond}},\ and\ \bibinfo {author} {\bibfnamefont {S.}~\bibnamefont
  {Boyd}},\ }\bibfield  {title} {\bibinfo {title} {A rewriting system for
  convex optimization problems},\ }\href@noop {} {\bibfield  {journal}
  {\bibinfo  {journal} {Journal of Control and Decision}\ }\textbf {\bibinfo
  {volume} {5}},\ \bibinfo {pages} {42} (\bibinfo {year} {2018})}\BibitemShut
  {NoStop}%
\end{thebibliography}%

\section{Appendix A: Methods for Data Acquisition and Processing}

In this appendix, we describe the methods used for data acquisition and processing. Two methods are introduced as follows.

\subsection{Method 1: Quantum state reconstruction}

The quantum computer mentioned in this paper is the SpinQ Gemini mini pro. In its measurements, each readout pulse can only provide the diagonal elements of the density matrix, which correspond to probabilities. To obtain the remaining elements of the matrix, it is necessary to perform rotation operations to retrieve all elements on the off-diagonal, thereby constructing the density matrix of the quantum state. In a single-qubit system, the following operations are required to construct the density matrix: I, X, and Y represent the identity operation, a 90° rotation around the x-axis, and a 90° rotation around the y-axis, respectively. For a two-qubit system, the required operations are II, IX, IY, XI, XX, XY, YI, YX, and YY\cite{lee,long1}. By converting the off-diagonal elements of the density matrix into diagonal elements using these operations, all elements can be determined by solving the equations. We illustrate this with a two-qubit example, where the matrices for the nine operations are:

\begin{align*}
II=
\begin{bmatrix}
   1 & 0 & 0 & 0 \\ 
   0 & 1 & 0 & 0 \\ 
   0 & 0 & 1 & 0 \\ 
   0 & 0 & 0 & 1
 \end{bmatrix},
\end{align*}

\begin{align*}
IX=
\begin{bmatrix}
   \sqrt{2}+0i & 0-\sqrt{2}i & 0+0i & 0+0i \\ 
   0-\sqrt{2}i & \sqrt{2}+0i & 0+0i & 0+0i \\ 
   0+0i & 0+0i & \sqrt{2}+0i & 0-\sqrt{2}i\\ 
   0+0i & 0+0i & 0-\sqrt{2}i & \sqrt{2}+0i
 \end{bmatrix},
\end{align*}

\begin{align*}
IY=
\begin{bmatrix}
   \sqrt{2} & -\sqrt{2} & 0 & 0 \\
   \sqrt{2} & \sqrt{2} & 0 & 0 \\ 
   0 & 0 & \sqrt{2} & -\sqrt{2} \\ 
   0 & 0 & \sqrt{2} & \sqrt{2}
 \end{bmatrix},
\end{align*}

\begin{align*}
XI=
\begin{bmatrix}
   \sqrt{2}+0i & 0+0i & 0-\sqrt{2}i & 0+0i \\
   0+0i & \sqrt{2}+0i & 0+0i & 0-\sqrt{2}i \\ 
   0-\sqrt{2}i & 0+0i & \sqrt{2}+0i & 0+0i \\ 
   0+0i & 0-\sqrt{2}i & 0+0i & \sqrt{2}+0i
 \end{bmatrix},
\end{align*}

\begin{align*}
XX=
\begin{bmatrix}
   \frac{1}{2}+0i & 0-\frac{1}{2}i & 0-\frac{1}{2}i & -\frac{1}{2}+0i \\ 
   0-\frac{1}{2}i & \frac{1}{2}+0i  & \frac{1}{2}+0i & 0-\frac{1}{2}i \\ 
   0-\frac{1}{2}i & -\frac{1}{2}+0i & \frac{1}{2}+0i  & 0-\frac{1}{2}i\\ 
   -\frac{1}{2}+0i & 0-\frac{1}{2}i & 0-\frac{1}{2}i & \frac{1}{2}+0i
 \end{bmatrix},
\end{align*}

\begin{align*}
XY=
\begin{bmatrix}
   \frac{1}{2}+0i & -\frac{1}{2}+0i & 0-\frac{1}{2}i & 0+\frac{1}{2}i \\ 
   \frac{1}{2}+0i & \frac{1}{2}+0i & 0-\frac{1}{2}i& 0-\frac{1}{2}i \\ 
   0-\frac{1}{2}i & 0+\frac{1}{2}i & \frac{1}{2}+0i & -\frac{1}{2}+0i \\ 
   0-\frac{1}{2}i & 0-\frac{1}{2}i & \frac{1}{2}+0i & \frac{1}{2}+0i
 \end{bmatrix},
\end{align*}

\begin{align*}
YI=
\begin{bmatrix}
   \sqrt{2} & 0 & -\sqrt{2} & 0 \\ 
   0 & \sqrt{2} & 0 & -\sqrt{2} \\ 
   \sqrt{2} & 0 & \sqrt{2} & 0 \\ 
   0 & \sqrt{2} & 0 & \sqrt{2}
 \end{bmatrix},
\end{align*}
 
\begin{align*}
YX=
\begin{bmatrix}
   \frac{1}{2}+0i & 0-\frac{1}{2}i & -\frac{1}{2}+0i & 0+\frac{1}{2}i \\ 
   0-\frac{1}{2}i & \frac{1}{2}+0i & 0+\frac{1}{2}i & -\frac{1}{2}+0i\\ 
   \frac{1}{2}+0i & 0-\frac{1}{2}i & \frac{1}{2}+0i & 0-\frac{1}{2}i \\ 
   0-\frac{1}{2}i & \frac{1}{2}+0i & 0-\frac{1}{2}i & \frac{1}{2}+0i
 \end{bmatrix},
\end{align*}

\begin{align*}
YY=
\begin{bmatrix}
   \frac{1}{2} & -\frac{1}{2} & -\frac{1}{2} & \frac{1}{2} \\ 
   \frac{1}{2} & \frac{1}{2} & -\frac{1}{2} & -\frac{1}{2} \\ 
   \frac{1}{2} & -\frac{1}{2} & \frac{1}{2} & -\frac{1}{2} \\ 
   \frac{1}{2} & \frac{1}{2} & \frac{1}{2} & \frac{1}{2}
 \end{bmatrix}.
\end{align*}

In a two-qubit system, \(\rho\) can be represented as a \(4 \times 4\) matrix, and the matrix form of \(\rho\) is given by equation \ref{density matrix1} and \ref{density matrix}. For all elements of \(\rho\), the quantum computer directly outputs signals that can only provide the elements \(x_1\), \(x_5\), \(x_8\), and \(x_{10}\), which correspond to \(\rho_{11}\), \(\rho_{22}\), \(\rho_{33}\), and \(\rho_{44}\), respectively. To obtain the other elements of the matrix, we must perform nine operations on the system to transform the required elements to the positions labeled as 11, 22, 33, and 44 in the density matrix, making them measurable. The readout provides the rotated matrix elements \(\rho_{11}^{'}\), \(\rho_{22}^{'}\), \(\rho_{33}^{'}\), and \(\rho_{44}^{'}\). These rotated matrix elements are linear combinations of the original matrix elements. Each operation yields four equations, and nine operations result in \(9 \times 4 = 36\) equations, with 16 unknowns. By solving this system of equations, the 16 unknowns can be determined, thus obtaining the complete matrix.

\subsection{Method 2: Convex optimization}

Due to noise in quantum computers, the measurement outcomes from each rotation operation often deviate from the linear combinations of the original density matrix elements, causing the reconstructed matrix to fail to meet the positive semi-definiteness requirement. Therefore, it is necessary to optimize the reconstructed matrix to ensure it becomes a valid density matrix.

The set of all quantum states generated by the circuit forms a convex set \cite{convex5}, which consists of Hermitian matrices with non-negative eigenvalues and unit trace. The fact that these matrices are both Hermitian and possess non-negative eigenvalues implies that the corresponding density matrix, describing a physical quantum state, is positive semi-definite, denoted as \(\rho \geq 0\). Consequently, the optimization problem can be formulated as a semidefinite program:

\begin{equation}
    \text{Minimize} \quad \|X - \rho\|^2 
    \label{eq:minimize},
\end{equation}

\begin{equation}
    \text{subject to} \quad \text{Tr}(X) = 1 
    \label{eq:trace_constraint},
\end{equation}

\begin{equation}
    \quad \text{and} \quad X \geq 0
    \label{eq:semidefinite_constraint}.
\end{equation}

In this formulation, \(X\) represents an Hermitian matrix variable, and \(\rho\) is the matrix reconstructed after the rotation operations. The objective function \(\|X - \rho\|^2\) in equation \ref{eq:minimize} is used to quantitatively capture the difference between \(X\) and \(\rho\), aiming to find the best Hermitian matrix \(X\) that satisfies the constraints: unit trace (\ref{eq:trace_constraint}) and positive semi-definiteness (\ref{eq:semidefinite_constraint}). These constraints not only ensure the physical validity of the density matrix but also serve as a regularization mechanism, making the optimization problem well-posed \cite{vogel1990constrained,kirsch2011introduction,souopgui2016incremental}.

Since the feasible solution set for \(\rho\) is convex and the cost function \(\|X - \rho\|^2\) is convex within this set, combined with the linear trace constraint, the optimization problem is inherently a convex optimization problem. Convex optimization problems are particularly advantageous because every local minimum is guaranteed to be a global minimum, ensuring that the problem has a unique optimal solution \cite{convex4}. 

This structured approach allows for efficient numerical methods to be applied, facilitating the practical implementation of state reconstruction procedures.

To solve this semidefinite programming problem, we use the open-source Python package CVXPY \cite{cvxpy1, cvxpy2, cvxpy3}, which provides efficient numerical methods for these types of problems, making the implementation of the state reconstruction process straightforward.

\end{document}